\documentclass[sigconf]{acmart}

\usepackage{subfigure}

\AtBeginDocument{%
  \providecommand\BibTeX{{%
    \normalfont B\kern-0.5em{\scshape i\kern-0.25em b}\kern-0.8em\TeX}}}

\copyrightyear{2024}
\acmYear{2024}
\setcopyright{acmlicensed}\acmConference[CHI '24]{Proceedings of the CHI Conference on Human Factors in Computing Systems}{May 11--16, 2024}{Honolulu, HI, USA}
\acmBooktitle{Proceedings of the CHI Conference on Human Factors in Computing Systems (CHI '24), May 11--16, 2024, Honolulu, HI, USA}
\acmDOI{10.1145/3613904.3642029}
\acmISBN{979-8-4007-0330-0/24/05}

\begin{document}

\title[Gesture-Based Interaction between Autonomous Vehicles and Pedestrians]{``It Must Be Gesturing Towards Me": Gesture-Based Interaction between Autonomous Vehicles and Pedestrians}

\author{Xiang Chang}
\authornote{Both authors contributed equally to this research.}
\orcid{0000-0003-1450-646X}
\affiliation{%
  \institution{Institute for AI Industry Research, Tsinghua University}
  \city{Beijing}
  \country{China}}
\email{eiuanin@163.com}

\author{Zihe Chen}
\authornotemark[1]
\orcid{0000-0002-9753-971X}
\affiliation{%
  \institution{Institute for AI Industry Research, Tsinghua University}
  \city{Beijing}
  \country{China}}
\email{948893227@qq.com}

\author{Xiaoyan Dong}
\orcid{0009-0000-0209-9575}
\affiliation{%
  \institution{Institute for AI Industry Research, Tsinghua University}
  \city{Beijing}
  \country{China}}
\email{dnxnyn99@163.com}

\author{Yuxin Cai}
\orcid{0009-0002-5406-2509}
\affiliation{%
  \institution{Institute for AI Industry Research, Tsinghua University}
  \city{Beijing}
  \country{China}}
\email{caiyx20@163.com}

\author{Tingmin Yan}
\orcid{0009-0008-3759-7978}
\affiliation{%
  \institution{Institute for AI Industry Research, Tsinghua University}
  \city{Beijing}
  \country{China}}
\email{275877004@qq.com}

\author{Haolin Cai}
\orcid{0009-0007-6587-0970}
\affiliation{%
  \institution{Institute for AI Industry Research, Tsinghua University}
  \city{Beijing}
  \country{China}}
\email{862422137@qq.com}

\author{Zherui Zhou}
\orcid{0009-0002-4047-1213}
\affiliation{%
  \institution{Institute for AI Industry Research, Tsinghua University}
  \city{Beijing}
  \country{China}}
\email{zherui.zhou@outlook.com}

\author{Guyue Zhou}
\orcid{0000-0002-3894-9858}
\affiliation{%
  \institution{Institute for AI Industry Research, Tsinghua University}
  \city{Beijing}
  \country{China}}
\email{zhouguyue@air.tsinghua.edu.cn}

\author{Jiangtao Gong}
\authornote{Corresponding Author}
\orcid{0000-0002-4310-1894}
\affiliation{%
  \institution{Institute for AI Industry Research, Tsinghua University}
  \city{Beijing}
  \country{China}}
\email{gongjiangtao2@gmail.com}


\begin{abstract}
  Interacting with pedestrians understandably and efficiently is one of the toughest challenges faced by autonomous vehicles (AVs) due to the limitations of current algorithms and external human-machine interfaces (eHMIs). In this paper, we design eHMIs based on gestures inspired by the most popular method of interaction between pedestrians and human drivers. Eight common gestures were selected to convey AVs' yielding or non-yielding intentions at uncontrolled crosswalks from previous literature. Through a VR experiment (N1 = 31) and a following online survey (N2 = 394), we discovered significant differences in the usability of gesture-based eHMIs compared to current eHMIs. Good gesture-based eHMIs increase the efficiency of pedestrian-AV interaction while ensuring safety. Poor gestures, however, cause misinterpretation. The underlying reasons were explored: ambiguity regarding the recipient of the signal and whether the gestures are precise, polite, and familiar to pedestrians. Based on this empirical evidence, we discuss potential opportunities and provide valuable insights into developing comprehensible gesture-based eHMIs in the future to support better interaction between AVs and other road users.
\end{abstract}

\begin{teaserfigure}
  \includegraphics[width=\textwidth]{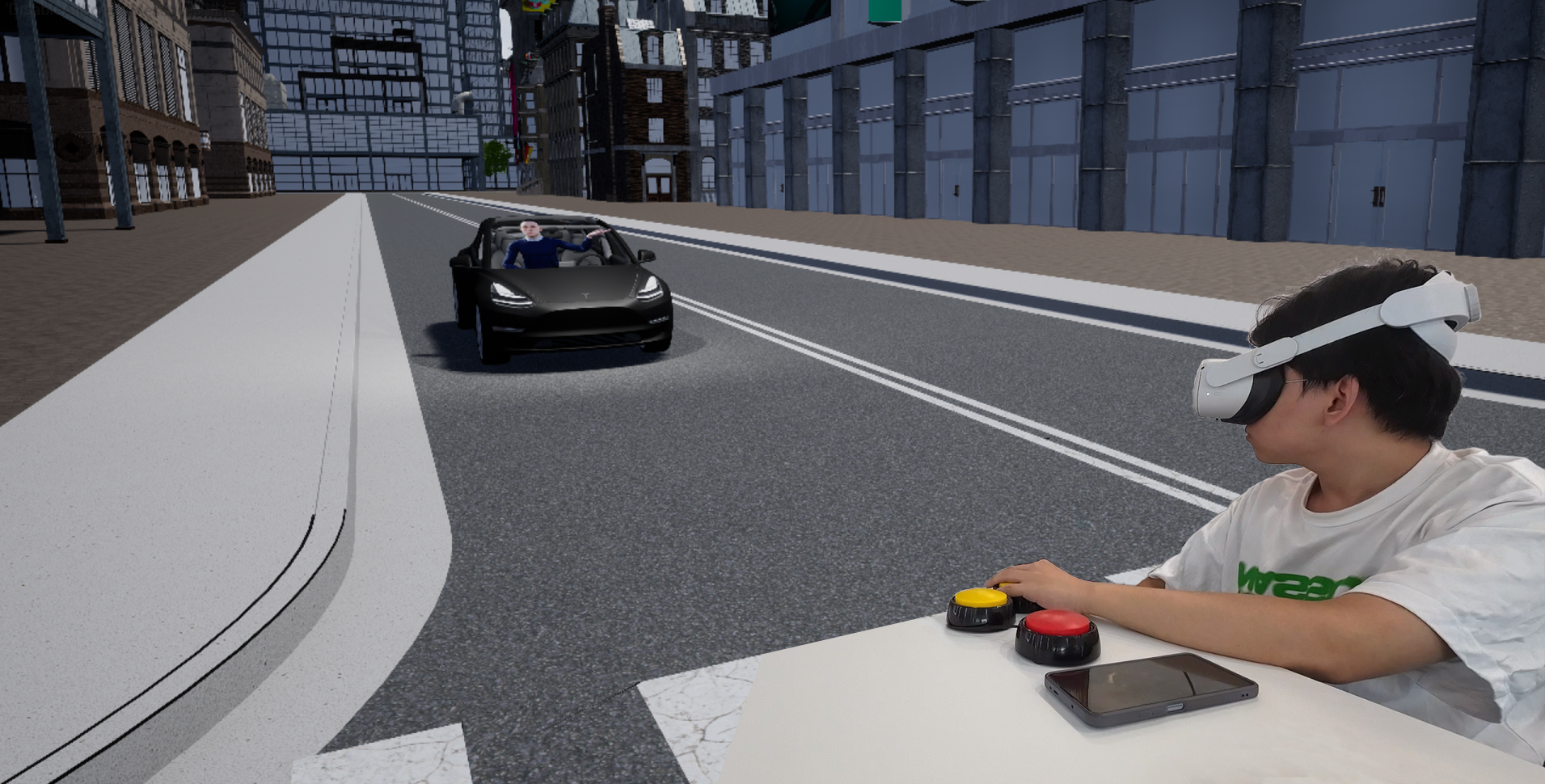}
  \caption{To evaluate the responses of pedestrians when facing different eHMI cars, we developed a virtual scenario of a pedestrian interacting with a self-driving automobile at an uncontrolled intersection. }
  \label{fig:teaser}
\end{teaserfigure}

\maketitle
\section{Introduction}
With the development of autonomous vehicles (AVs), having the ability to effectively communicate and interact with other road users in traffic is challenging for AVs~\cite{brooks2017big}.
In a scenario where an AV comes across a pedestrian who intends to cross the road at an uncontrolled crossing, the determination of priority is not clearly defined. Due to safety and regulatory considerations, pedestrians are typically regarded as obstacles in the majority of today's AV systems, prompting cars to stop when encountering them.
Consequently, pedestrians may exploit this characteristic of AVs and gain an advantage over them in all interactions, including intentionally placing themselves in front of AV to impede their forward motion~\cite{Mengdi}. 
Circumstances that pedestrians hinder the AVs from making any significant advancements were commonly referred to as the "Frozen Robot Problem"~\cite{5654369}. The widespread occurrence of this situation will cause severe traffic chaos. 

In such a scenario, AVs must possess the capability of not yielding to pedestrians and other participants in road traffic~\cite{doi:10.1177/0739456X16675674}.
Considering the safety of pedestrians, pedestrians need to obtain prior notification of the non-yielding intentions of AVs before the cars pass in front of them.
Hence, the capacity of AVs to effectively convey their intentions and communicate with pedestrians is beneficial. 
To this end, external human-machine interfaces (eHMIs) 
were introduced into AVs to provide the capability to convey important information to external traffic participants in their surroundings~\cite{moore_case_2019,lee_learning_2022,dietrich_projection-based_2018}. 
Nevertheless, the effective implementation of interaction between autonomous vehicles and other road users, particularly pedestrians and cyclists, is still worth worrying about since other road users may not even notice that they are involved in interaction with an autonomous vehicle~\cite{doi:10.1177/0739456X16675674}.

\begin{figure}[h]
    \centering
    \includegraphics[width=0.8\linewidth]{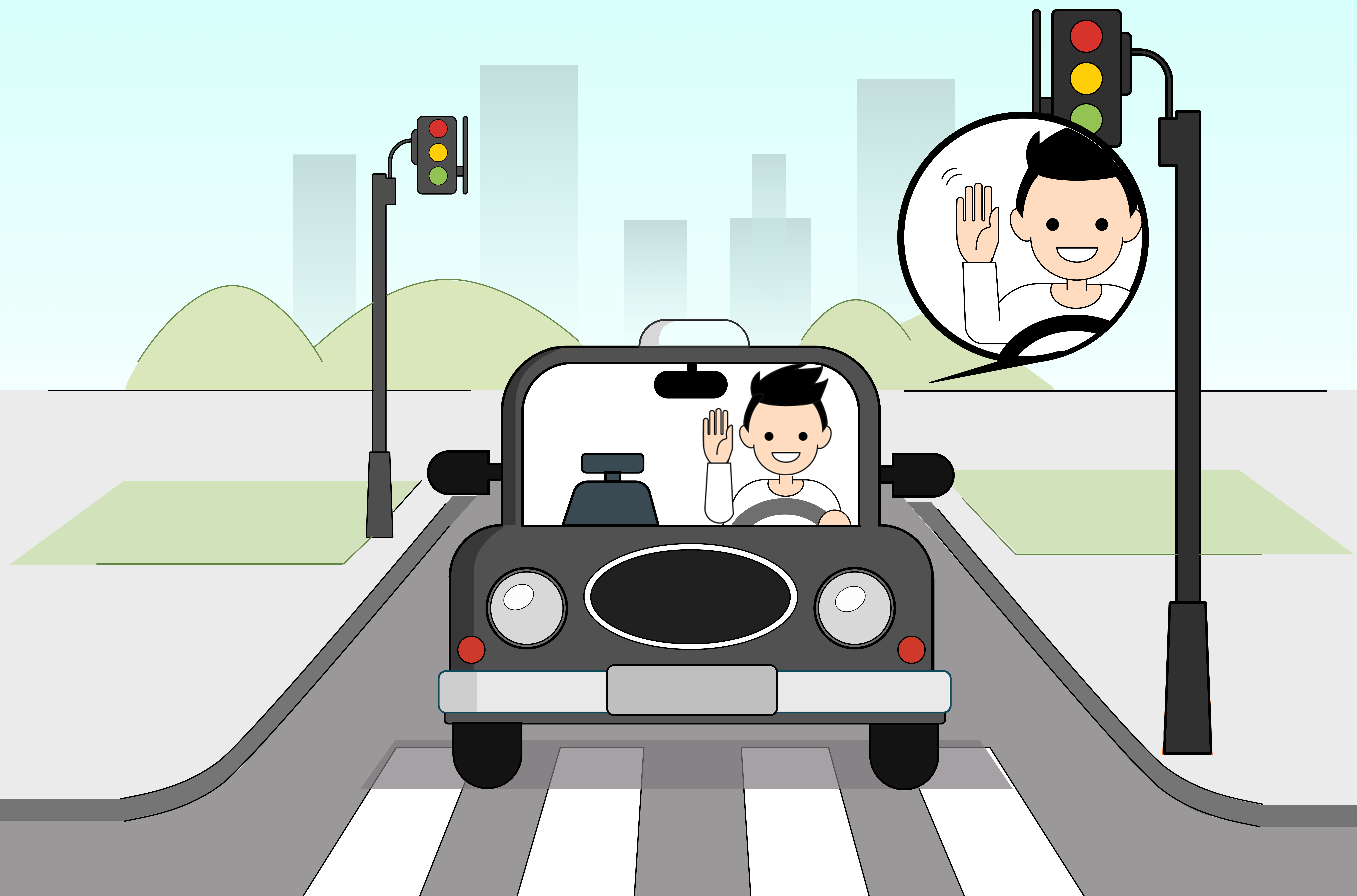}
    \caption{Pedestrian-driver interaction using gesture at uncontrolled crosswalks.}
    \label{fig:Intro-cross}
\end{figure}

Gestures, which are frequently employed as a means of communication between drivers and pedestrians or other vulnerable road users~\cite{8241847} (See Fig.\ref{fig:Intro-cross}), have received comparatively less attention as cues to be referred to in existing designs and research on eHMI, especially compared with lights and texts. 
However, the previous studies conducted on gesture-related interactions between pedestrians and autonomous vehicles have mostly focused on the AV's detection of pedestrian intents via identifying and understanding their gestures~\cite{epke_i_2021}.
The main objective of this study is to fill in the existing research gap about gesture-based eHMI to facilitate interactions between AVs and pedestrians.
We aimed to compare the relative effectiveness of gesture-based eHMIs with other eHMIs. Additionally, the study seeks to identify the most suitable gestures from a wide range of popular gestures to provide valuable insights for the design of future eHMIs designed for AVs.
Thus, our research questions are listed as follows:

\begin{itemize}
    \item [RQ1:] What are the differences between the effects of eHMIs? Which gestures work best?
    \item[RQ2:] What underlying factors contribute to this difference?
\end{itemize}

To answer the above questions, we conducted a virtual reality experiment (N1 = 31) and a following online survey (N2 = 394) to validate the generalizability of the results. 
Firstly, in our virtual scenario, we asked participants to cross the road as fast as possible while ensuring their safety to examine the effects of eight gesture-based eHMIs (four yielding and four non-yielding gestures) developed by us compared with two SOTA eHMIs (one indicating yielding intent and the other indicating no yielding) and AVs without any eHMI. 
Due to the limited acceptance of VR experiments among certain demographics, such as predominantly young people, to obtain a more comprehensive variety of perspectives from the wider population, we developed a  questionnaire involving 394 participants covering road users of all age groups. The questionnaire further checks the three factors for assessing gesture-based eHMIs, including clarity, familiarity, and politeness. 

To summarize, our work makes the following contributions:
i) a gesture-based eHMI concept design inspired by human driver---pedestrian gesture interaction; ii) a gesture set for driver-pedestrian interaction selected from multiple transportation scenarios and literature; iii) a controlled study with 31 participants to investigate differences among gestures' performance in a VR simulation experiment; iv) a wider population survey with 394 participants to further validate the efficiency of each gesture. Our research provides solid empirical evidence and in-depth design implications for designing gesture-based eHMI for AV in the future.

\section{RELATED WORK}
\subsection{Autonomous Vehicles-Pedestrian Interaction Design}
Pedestrians are one of the most vulnerable users on the road. Misjudgments between vehicles and pedestrians can lead to accidents and even cause casualties~\cite{vanlaar2016fatal}. Compared to traditional cars, autonomous vehicles lack certain methods of communication~\cite{colley2020unveiling}. Therefore, the interaction between autonomous vehicles and pedestrians will affect pedestrian safety~\cite{el2020pedestrian}. 
The interaction issues between AVs and pedestrians have attracted a lot of attention in HCI research.
Past research has explored various design approaches to facilitate AV-pedestrian interaction, such as audio signals (engine sounds, bell rings, and soft alarm sounds)~\cite{zhang2020user}, Light Signals~\cite{dey_color_2020, dey2022investigating,fridman_walk_2017}, Augmented Reality (AR) projection on the road~\cite{tabone2021towards}, Mobile Applications and Connected Devices~\cite{walch2020crosswalk,hollander2020save}, Internet Vehicle-to-Everything (V2X) Technology~\cite{zoghlami20235g}, Interfaces on Street Infrastructure~\cite{mahadevan2018communicating} and external Human-Machine interface (eHMI)~\cite{faas2020longitudinal,liu2023pre}.

Empirical research on the aforementioned methods indicates that these approaches have their respective advantages in terms of clarity, familiarity, and politeness, yet they still present challenges.
Regarding \textbf{clarity}, the effectiveness of Audio Signals is limited in conveying risk levels~\cite{zhang2020user}; Light Signals are also inefficient in ambiguous situations and busy scenarios~\cite{dey_color_2020, dey2022investigating,fridman_walk_2017}; Street Infrastructure Interfaces may cause visual confusion and fail to indicate multiple pedestrians' intents, affecting clarity~\cite{mahadevan2018communicating}.
Regarding \textbf{familiarity}, AR~\cite{tabone2021towards} offers novel interaction methods but its complexity can increase cognitive load, reducing familiarity; the effectiveness of Audio Signals in conveying risk levels varies across environments, limiting clarity; Mobile Apps and Connected Devices provide intuitive, personalized interactions for understanding AVs' intentions but their effectiveness is contingent on widespread adoption and user familiarity~\cite{walch2020crosswalk,hollander2020save}; V2X Technology enhances safety through effective communication but faces challenges in network latency and reliability, reflecting limited user familiarity~\cite{zoghlami20235g}.
Regarding \textbf{politeness}, Audio Signals may be perceived as impolite, especially in varied environments where they can be intrusive~\cite{zhang2020user}; Light Signals lack politeness in multi-vehicle scenarios, leading to visual confusion~\cite{dey_color_2020, dey2022investigating,fridman_walk_2017}; AR Technology may potentially be more polite by personalizing interaction information, minimizing pedestrian disturbance and misunderstanding~\cite{tabone2021towards}.

From the studies mentioned above, it is evident that an effective pedestrian-AV interaction design should be taken its clarity, familiarity, and politeness into consideration. Therefore, in this paper, we also comprehensively take into account these factors for the selection and evaluation of gestures.

\subsection{The State of the Art (SOTA) Dispaly-based eHMI.}

Display-based eHMI has become a mainstream research direction. Display-based eHMI uses visual screens or projections that can display dynamic information. These display technologies have seen significant development in recent years and are being integrated into automobiles. For example, AUO showcased its Micro LED automotive display technology at CES 2023. This technology offers high brightness, high contrast, a wide color gamut, and rapid response times. The PML programmable smart headlights and ISD intelligent interaction lights equipped in HiPhi X can project interactions with pedestrians. These display technologies can be integrated inside vehicles, such as on windshields or the body of the car, or even projected onto the road or surrounding environment. 
In recent years, a large number of display-based eHMI design concepts have been proposed or applied. Examples include the Mercedes-Benz F015 concept car (Daimler, 2015)\cite{benz2015mercedes}, the Nissan IDS concept car (Nissan Motor Corporation, 2015), the Volvo Concept 360 (Volvo Cars, 2018), the Smart EQ ForTwo concept (Daimler, 2017)\cite{daimler2017autonomous}, the BMW VISION NEXT 100 (BMW, 2016), and the Jaguar/Land Rover Virtual Eyes concept (Jaguar Land Rover, 2018) \cite{rover2018virtual}.

In addition to industrial products, the academic community has also conducted empirical research on various display-based eHMIs. Bazilinskyy et al.~\cite{bazilinskyy2019survey} found textual eHMIs in AVs clearer than non-textual ones, especially with zebra crossings, and recommended egocentric text for safety. Werner et al.~\cite{werner2018new} identified turquoise as the best color for AV external lighting, emphasizing visibility and uniqueness. Lau et al.~\cite{lau2022toward} showed that dynamic eHMI enhances safety in yielding AVs but warned against mismatches between eHMI signals and vehicle actions. Dey et al.~\cite{dey2020color} discovered cyan with flashing or pulsing animations as effective for eHMIs in AVs, aiding pedestrian interaction.

While there has been considerable research on eHMI, it has been somewhat limited to interaction designs already present in traditional vehicles. Therefore, in this paper, we integrate the SOTA eHMI design and studies into an optimally effective eHMI to be used as a control group design (see details in Section~\ref{SubSec:Independent_Variables}, Fig.~\ref{fig:12CAR}). 

\subsection{Driver Gestures and eHMI}

In Katz et al.'s~\cite{katz1975experimental} study on driver-pedestrian interaction in cross-conflicts, it was found that after vehicle-pedestrian conflicts, drivers and pedestrians often communicate through non-verbal means like gestures and eye contact. Social norms or tacit communication play a significant role in predicting each other's intentions
~\cite{nilsson2015action}. For instance, eye contact, gestures between pedestrians and drivers, or changes in vehicle speed can convey intentions
~\cite{ackermann2019deceleration}. Stanciu et al.~\cite{stanciu2018pedestrians} summarized interactions between vehicles and pedestrians using gestures, eye contact, or headlights. Drivers use gestures to encourage other road users to pass or stop~\cite{vsucha2014road}. Pedestrians also use gestures to communicate with vehicles~\cite{gruenefeld2019vroad}. Research shows that gestures are a common method of traffic communication, improving driver-pedestrian interaction~\cite{crowley2011effects,zhuang2014pedestrian}. Gestures (like waving) are common practices in negotiating right-of-way between pedestrians and drivers~\cite{dey2017pedestrian}. Similarly, eye contact is considered a social cue; it's an indicative signal conveying to whom one is speaking and that subsequent actions or information will be meaningful~\cite{bockler2014catching}. Šucha~\cite{vsucha2014road} confirmed that both pedestrians and drivers use eye contact for communication, with pedestrians using it to encourage drivers to yield~\cite{zhang2020user}.

Fass et al.~\cite{faas2020longitudinal,m2021calibrating} noted that with the increase of automated function vehicles, future traffic systems will be shared by autonomous and manually driven vehicles. In such scenarios, traditional communication methods between pedestrians and AVs, like eye contact and gestures, may no longer be effective~\cite{habibovic2018communicating,ackermann2019experimental}. Hence, many studies have analyzed pedestrian-driver interactions at crosswalks in urban settings and how these interactions influence the behavioral design of future AVs. The focus is on understanding interactions at crossroads between drivers and pedestrians, deriving insights from these interactions to guide the development of autonomous driving functions~\cite{mahadevan2018communicating,schneemann2016analyzing,sucha2017pedestrian,carmona2021ehmi}

Ackermann et al.~\cite{ackermann2019deceleration} believed that using existing communication methods (gestures, eye contact) by autonomous vehicles might be beneficial. These traditional and common behavior patterns of drivers expressing priority to pedestrians have been discussed between autonomous vehicles and pedestrians~\cite{zhanguzhinova2023communication}. Risto et al.~\cite{risto2017human} emphasized the need for AV developers to recognize and understand the impact and importance of motion gestures in daily traffic interactions, and to design effective and scalable forms of human-vehicle communication in multi-party interactions. A simple gesture action performed well in comparisons across different interfaces, as it clearly conveyed intentions~\cite{mahadevan2018communicating}. When gestures align with the information provided by eHMI, pedestrians' willingness to cross the road in front of autonomous vehicles operating in yielding mode will increase. Conversely, it will decrease when gestures and eHMI information are inconsistent~\cite{colley2022user}.

\begin{figure*}[h]
    \centering
    \includegraphics[width=0.95\linewidth]{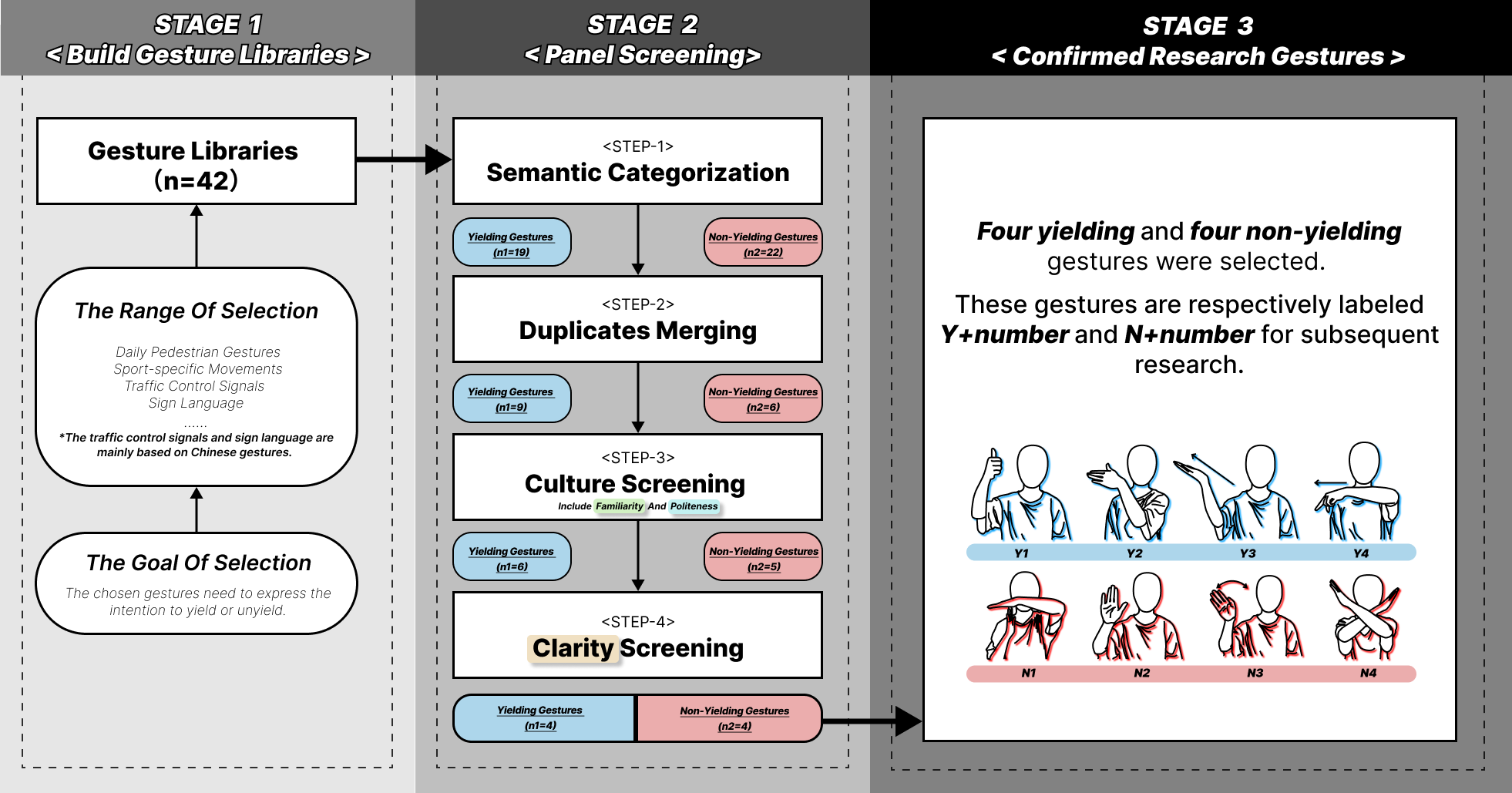}
    \caption{The Process of Gesture Selection, including three stages}
    \label{fig:Gesture Selection}
\end{figure*}

\section{Gesture Selection}
~\label{Sec:Gesture_Selection}
To assess the practicality of integrating gesture interaction into the eHMI design, researchers carried out a process of selecting suitable gestures. This involved three primary stages (Fig.\ref{fig:Gesture Selection}). In the initial stage, the researcher gathered gestures and created a gesture library. Depending on the needs of the study, the chosen gestures need to express the intention to yield or unyield. The selected gestures comprise daily pedestrian gestures, sport-specific movements, traffic control signals, and sign language~\cite{crowley2011effects, zhuang2014pedestrian, farber2016communication, weber2019crossing}. It should be noted that this research is taking place in China, so the traffic control signals and sign language are mainly based on Chinese gestures.

In the second stage, the gestures in the gesture library were evaluated and screened by a panel of six researchers. Step 1 was to categorize the gestures semantically. Gestures were classified based on their meanings, with affirming and passage-signaling gestures falling under the yielding category, while negation and passage-stop signaling gestures were categorized as non-yielding. Step 2 was to perform a duplicate merge. Similar gestures in key areas were combined and categorized together. Step 3 was to conduct a cultural screening. The study evaluated familiarity of gestures  and eliminated those that were deemed too culturally specific. In addition, the politeness of gestures was considered, and those with offensive meanings in certain cultures were excluded. For instance, the clasping gesture was removed because it is too Chinese, while the ``OK'' gesture was omitted due to its offensive implications in certain parts of Italy\cite{farber2016communication}. Step 4 was to screen for clarity of gestures. Gestures with too little amplitude and too much detail were screened out. See the supplementary material for sources and screenings of specific gestures.

\begin{figure*}[htb]
    \centering
    \includegraphics[width=0.85\linewidth]{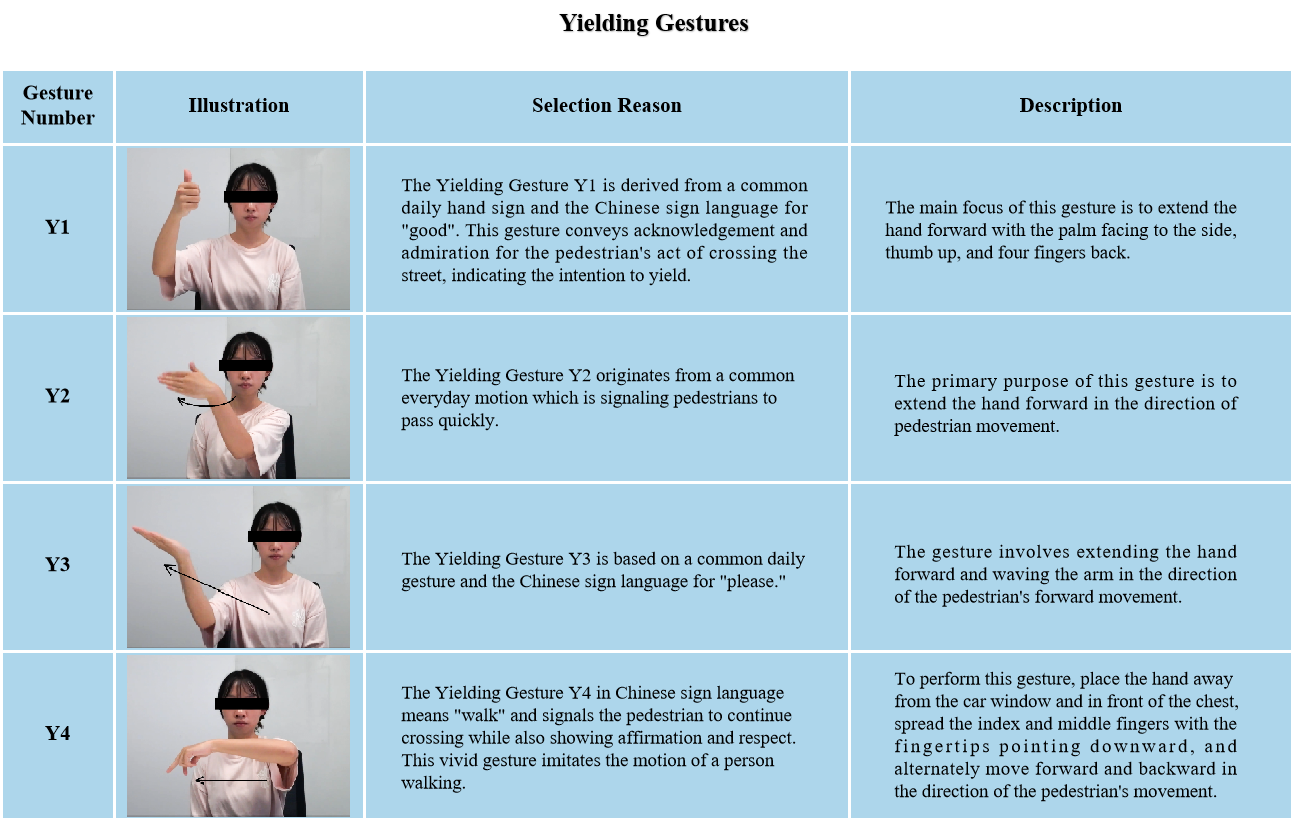}
    \caption{Four yielding gestures}
    \label{fig:YieldingGesture}
\end{figure*}

\begin{figure*}[htb]
    \centering
    \includegraphics[width=0.85\linewidth]{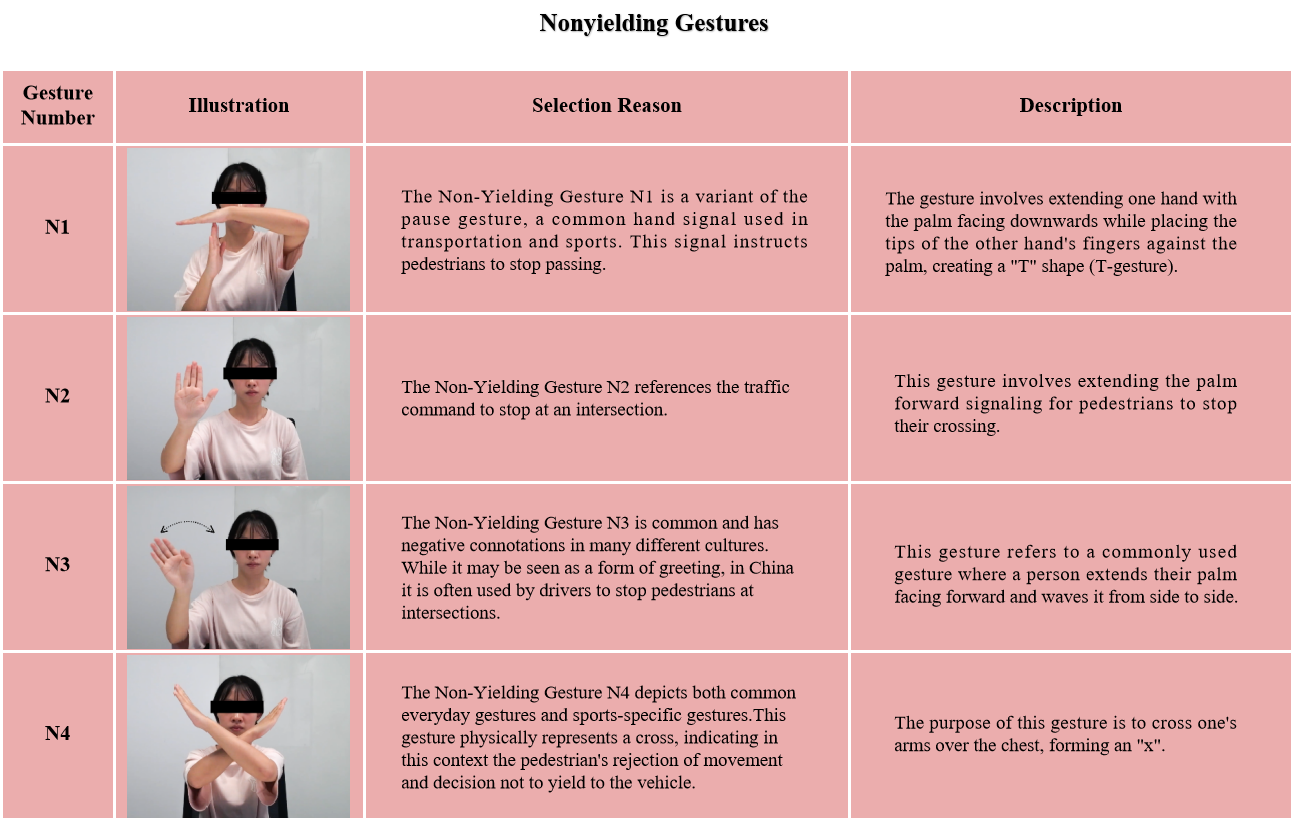}
    \caption{Four non-yielding gestures}
    \label{fig:NonyieldingGesture}
\end{figure*}

After the panel screening, four yielding gestures (Fig.~\ref{fig:YieldingGesture}) and four non-yielding gestures (Fig.~\ref{fig:NonyieldingGesture}) were selected. These gestures were respectively labeled Y+number and N+number for subsequent research.


\section{STUDY 1:  Virtual Reality Experiment}

Given the current limitations of AV technology and potential hazards associated with conducting tests involving AVs in real-world scenarios, virtual reality (VR) technology was widely utilized in human-AV interaction experiments to simulate and examine pedestrian-AV interactions to ensure safety and improve user comfort~\cite{mahadevan2019av,hollander2019investigating,deb2020communicating}.
However, VR experiments also have some limitations. For example, participants' perceptions in VR usually differ from their perceptions in reality, and complex operations might cause discomfort and symptoms such as dizziness and nausea, which may limit the range of our participants~\cite{oh2021feasibility,tian2022review,Blascovich_Loomis_Beall_Swinth_Hoyt_Bailenson_2002}.
Thus, in our research, we first conducted a VR experiment with careful design. Then we conducted an online survey to make sure of the generalization of our findings.

\subsection{Scenario and Task Design}

To explore the usability and performance of the selected gestures in facilitating pedestrian-AV interaction, we designed a VR experiment to compare all the gesture-based eHMIs along with baseline conditions.
The VR experiment took place in a virtual scenario (See Fig.~\ref{fig:VR-PROCESS}) where pedestrians and AVs met at uncontrolled crosswalks.
The participant's task is to control the virtual pedestrian's behaviors (gaze and move) in response to AVs with different eHMIs.
To address possible dizziness and nausea problems in pilot studies, we limited participants to sitting down and pressing two buttons that simulate walking in the virtual environment.

Before crossing, pedestrians usually observe vehicles around them and consider possible dangers~\cite{kotseruba2020they}.
We hypothesize that an effective eHMI should help reduce the amount of time pedestrians observe and hesitate. These times serve as indicators of the effectiveness of eHMI.\cite{dey2017pedestrian}

\begin{figure*}[h]
    \centering
    \includegraphics[width=1\linewidth]{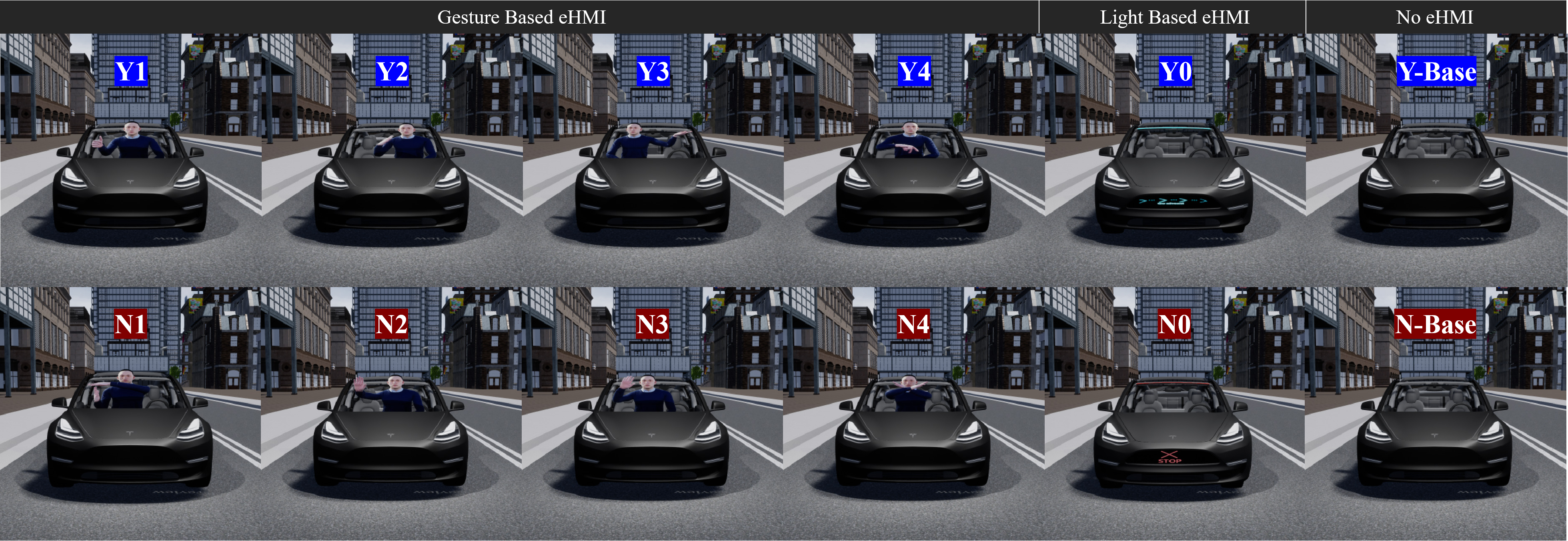}
    \caption{The implementation of each experiment conditions.}
    \label{fig:12CAR}
\end{figure*}

Thus, a within-subjects design was utilized in this study. The usability of 12 eHMI designs was assessed, including eight gesture-based eHMIs (four yielding and four non-yielding gestures) , two SOTA eHMIs (light-based, one yielding and one non-yielding) and two AVs without any eHMI (one yielding and one non-yielding), as shown in Fig.\ref{fig:12CAR}.
Each eHMI was presented to participants three times, and the sequence in which eHMI designs appeared was randomized. This randomization aimed to prevent subjects from discerning trends that could potentially influence the experimental outcomes. We have designed the system to randomly generate different types of AVs at each intersection. Each AV comes with its own eHMI and corresponding motion logic, all of which are programmed into an AV blueprint. In this experiment, there are 12 different blueprints. To ensure that each AV appears exactly 3 times, we will create two duplicates of each AV's blueprint, expanding the blueprint library to $12\times 3 = 36$ in total and assigning each blueprint a unique number from 1 to 36. As the pedestrian approaches the trigger point before an intersection, the program will randomly select an integer from this range and summon the corresponding AV. To ensure that each intersection corresponds to a unique blueprint without repetition, the system will exclude numbers that have already been randomly selected when moving to the next intersection.

Additionally, to encourage pedestrians to actively interpret eHMI signals and take action, a incentive mechanism is implemented in the experiment. Under conditions where the AV yields, a reward is triggered if the pedestrian's hesitation time is less than 5 seconds. Conversely, when encountering a non-yielding AV, a reward is earned by successfully navigating the pedestrian through. No additional reward is given for failing to meet these conditions. Each time an additional reward is triggered, participants will receive extra money on top of base compensation.



\subsection{Independent Variables}~\label{SubSec:Independent_Variables}

\textbf{eHMI Designs}: We assessed the 12 eHMIs designs. The SOTA eHMI design were labeled as Y0 and N0. AVs without any eHMI design were labeled as Y-base and N-base. The eHMI designs are summarized in Fig.\ref{fig:12CAR}.

\textbf{Encounter Times}: Each design was examined by each participant three times.

\subsection{Dependent Variables}~\label{Sec:Depenedent_variables}
The following measures were recorded or calculated to assess the effect of eHMIs:

\textbf{Error Rate:}
Whether the subject misjudged the intent of the AV to calculate the Error Rate, the understanding of the intentions of AVs is reported by the participant orally. For non-yielding AVs, if the pedestrian makes an error in judgment and walks up to the crosswalk without waiting for the car to pass, he or she is bound to be hit by the car. The occurrence of a collision is recorded. The error rate for an eHMI design was calculated as the number of times that its intention was misunderstood, divided by the total times that the eHMI design was examined by all participants.

\textbf{Difficulty in Comprehension:} The score of the difficulty of understanding the shown eHMI on a Likert-type scale of 1 to 5, reported by the participant orally.

\textbf{Perception of Danger:} The score of the perceived level of danger when passed by  AV with specific eHMI, on a Likert-type scale of 1 to 5, reported by the participant orally.

\textbf{Duration of Observation:} The time the participant spent observing the AV. Given that AVs always approach pedestrians from the left, pedestrians must turn their heads toward the left to observe the oncoming vehicle. In the VR, the camera's field of view is 90°, therefore the AV only fully enters the pedestrians’ field of view when they turn their head 50° to the left. Once the observation is complete, pedestrians should then redirect their attention back to the road surface in front of them. The duration of observation could be recognized as the time when they turned their heads more than 50° to the left.

\textbf{Duration of Hesitation:} The time the participant spent hesitating to go forward. During the experiment, the participant was given a yellow button to press and hold to control walking. The time that the participant did not press the button while crossing the intersection was recorded to be the measurement.

\textbf{Physiological Indicators:} Electrodermal activity(EDA), heart rate (HR), and heart rate variability (HRV), measured by wearable monitoring device——Empatica E4 Wristband. Physiological indicators were derived :

\emph{SCR\_Peaks\_N:} The number of peaks of the Phasic Skin Conductance Response (SCR).\cite{braithwaite2013guide}

\emph{HR:} The number of times the heart beats per minute.

\emph{HRV\_RMSSD}: The root mean squared differences of successive difference of intervals.\cite{salahuddin2007ultra,pham2021heart}

\subsection{Apparatus}
Using the Carla version of Unreal Engine\cite{Dosovitskiy17}, we designed a city scenery that closely resembles the real world. In order to simulate the real world as much as we could, we added streets, complex supermarkets, office buildings, residential architecture and plants, etc. The city street scenery is made by the software RoadRunner. As the dual carriageway, each lane is 5 meters wide, and both sides of the road have a sidewalk 3 meters wide.

\begin{figure}[h]
    \centering
    \includegraphics[width=1\linewidth]{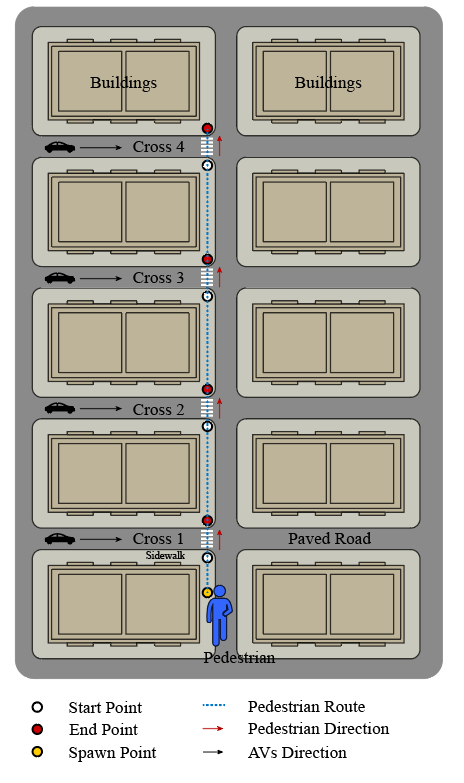}
    \caption{The implementation of each experiment conditions.}
    \label{fig:VR-Layout}
\end{figure}

\begin{figure}[h]
    \centering
    \includegraphics[width=1\linewidth]{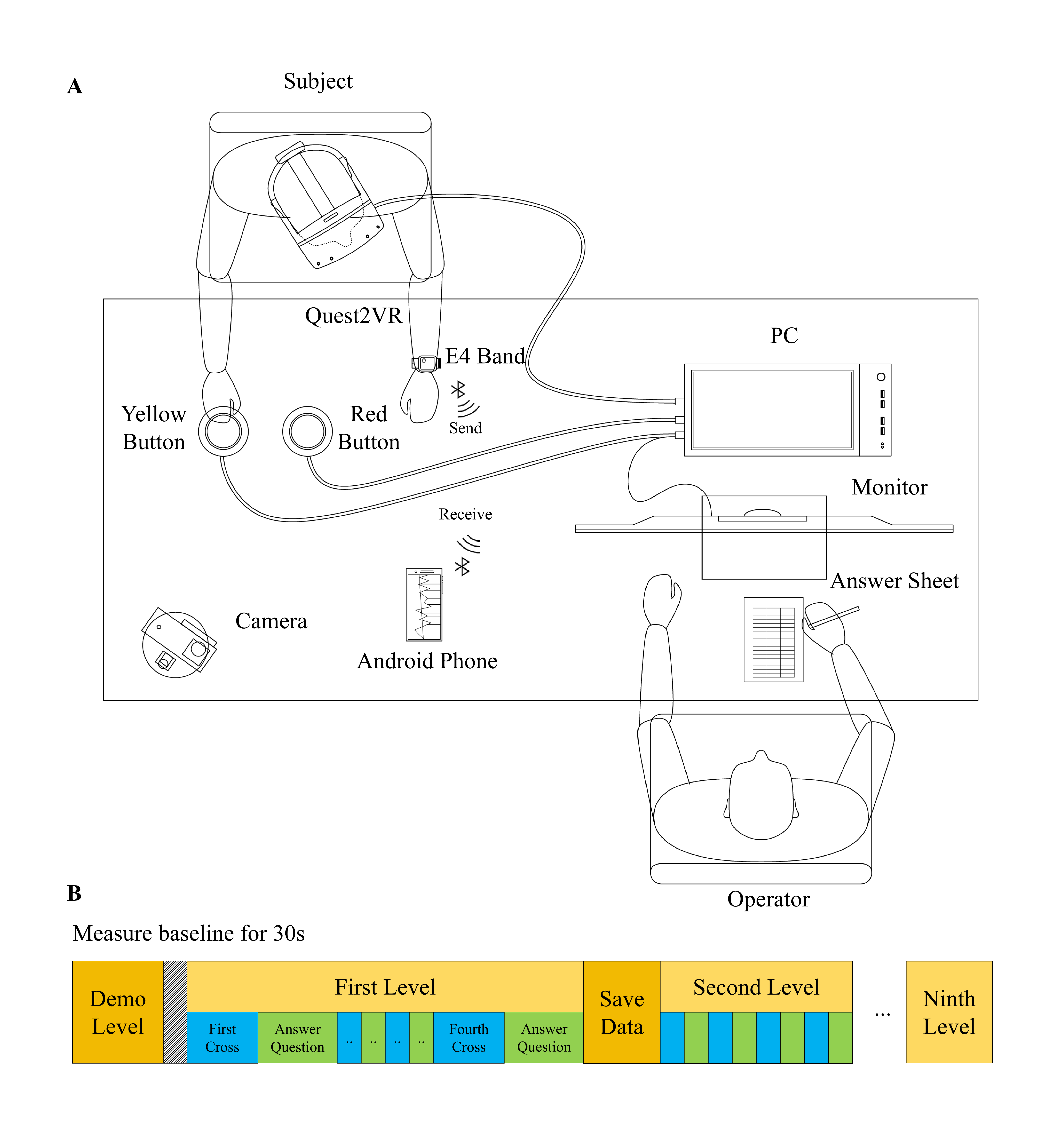}
    \caption{Layout of the experiment environment and experiment procedure.}
    \label{fig:Procedure}
\end{figure}

Pedestrians will encounter AVs at crossroads. There are zebra crossings at the intersection, but there are no traffic lights. Pedestrians need to cross every crossroad. For a crossing, a starting point is marked 5 meters in front of the zebra crossing, and an ending point is marked at the end of the zebra crossing. Between the starting point and the ending point is the experimental stage, where pedestrians understand the intention of the autonomous vehicle and cross the road. The layout of the scenario in the virtual environment and the routes of AVs and pedestrians are illustrated in Fig.\ref{fig:VR-Layout}.

The generated positions and behaviors of AVs are standardized. We meticulously designed the motion behaviors and transition points of AVs, making it challenging for participants to infer vehicle intent through implicit information. The AV's driving behavior consists of three stages: the first stage involves the AV traveling at a speed of 43.2 km/h for 2 seconds after being generated, during which the eHMI signals on the vehicle start playing. In the second stage, the AV undergoes a 2-second deceleration, reduces its speed to 14.4 km/h, and continues traveling for 9 seconds. In the third stage, the yielding and non-yielding AVs exhibit different motion behaviors. If the current AV decides not to yield, it continues at a constant speed of 14.4 km/h through the intersection. If the current AV decides to yield, it decelerates to a stop 4 meters before the crosswalk.


According to Fig.\ref{fig:Procedure} A, the experiments are conducted in an empty indoor environment in which tables and chairs are placed.  Beyond the desk, there is a PC (CPU: intel i5 13600kf, GPU: Nvidia rtx3080 10GB, RAM:32GB, ROM:4TB) with a virtual environment, an android cellphone ( XIAOMI MIX 2s, CPU: snapdragon845, ROM:6GB),  a physiological state monitoring wristband (Empatica E4 Wristband), a VR headset (Oculus Quest2), a display device(34-inch display or Redmi) and two physical buttons. The VR device was connected to the PC with a USB 3.0 cable. The two buttons connected to the computer are red and yellow, separately. While running, the PC will sync the picture to the monitor, and the E4 wristband is connected to an Android cellphone by Bluetooth. A paper answer sheet was used to record how participants rated their responses during the experiment.

\subsection{Procedure}
After welcoming the participants and introducing the research target briefly, we asked them to sign an informed consent form. Next, we explained experimental tasks and procedures in detail and told them how many buttons they could use and the data needed to be collected. The experimenter wore a VR headset to demonstrate the approaches to operating for participants. Before continuing, we ensured that we had solved any unclear problems posted by participants. Afterward, the researchers helped the participants wear E4 wristbands and launched the acquisition software running on the Android cellphone. Then, the participants were helped to put on the VR headset and adjust it to a position where it was comfortable to wear. Until confirming there were no problems, participants entered the virtual environment of the first round. To collect baseline physiological data for 30 seconds, participants were also asked to breathe steadily without any movement. 30 seconds later, the researcher issued a start command, and the participants took over the operation buttons to start the formal experiment. 

\begin{figure*}[h]
    \centering
    \includegraphics[width=1\linewidth]{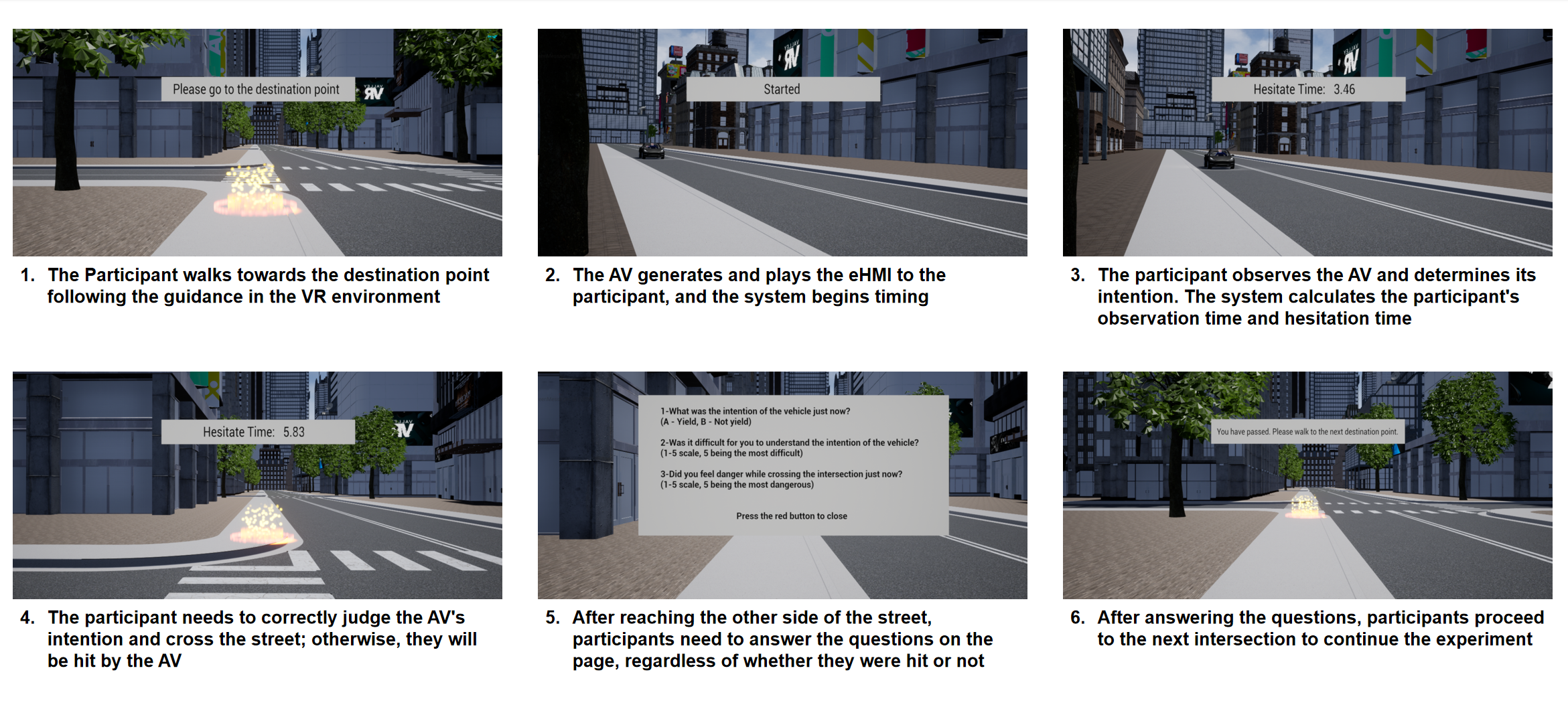}
    \caption{The procedure of the VR experiment.}
    \label{fig:VR-PROCESS}
\end{figure*}

During a round of the experiment, participants pressed and held the yellow button to control to walk to the starting point according to the instructions in VR. When the pedestrian in VR played by the participant arrived at the starting point of each crossroad, Unreal Engine started to record all the data required. In the meantime, AV was generated and approaching the crossroad by the left side of the pedestrian. Participants need to judge the vehicle intention according to the eHMI information designed in the experiment and decide whether to go before the AV. When pedestrians arrive at the ending point of the crossroad, a crossing is considered to be completed. Participants need to answer the questions at the pop-up windows orally. Below is the content of the question:

\begin{itemize}
    \item [1.] What is the intention of the vehicle? ( A. Yield B. Not to yield)
    \item [2.] How difficult is it for you to understand the intentions of AVs? (1-5 points, with five being the most difficult)
    \item [3.] How dangerous do you feel when you cross the road? (1-5 points, with five being the most dangerous)
\end{itemize}


The researcher recorded the answers in writing. Participants can press the red button to close the pop-up window after finishing the questions and then use the yellow button to control the pedestrian to go to the next intersection. The procedure of the experiment is illustrated in Fig.\ref{fig:VR-PROCESS}.

Semi-structured interviews lasting no more than 30 minutes were conducted after the experiments. Participants need to review the process of the experiment together and the questions answers. If they were bumped/ got a wrong answer to question 1 in the experiment, they need to check the reason. During the process of answering, participants can view the video to review different kinds of gestures anytime. For questions 2 and 3, after review, participants need to re-evaluate and rank all different yielding or non-yielding eHMI effects and explain why. Finally, there were a few questions related to AVs and gesture-based eHMI at the end. 

\subsection{Participants}
In August 2023, this study recruited adults between 18 and 55 years old in Beijing. The recruitment criteria were as follows:
i) No cognitive disorder, basic social cognition, and judgment ability, normal mental state. ii) No perception disorders, visual, auditory, and sensory functions are normal. iii) Normal athletic ability without any physical disability. iv) No eye or skin infections. v) Participants have long-term residency in the North China region, being familiar with local traffic rules and cultural norms. vi) Participants have previous experience with VR devices and have not exhibited any adverse reactions. vii) Participants are in a normal state and have abstained from alcohol or similar substances before participating in the tests.

Finally, we selected 33 healthy adults with VR play experience from many applicants to participate in this experiment, and all the participants needed to voluntarily sign informed consent before the experiment. Participants played as pedestrians to interact with automatic vehicles in the VR world. Each participant received 60 RMB as compensation after the experiment finished. This experiment has been approved by the local university’s Institutional Review Board (IRB).

After being scrutinized, the data from 31 participants was considered reliable and analyzed. Due to equipment problems during the experiments with two participants, the data for them were incomplete. Consequently, the data from these two participants were excluded from the analysis.

\subsection{Analysis}            
We checked whether participants' reported interpretations of the AVs' intentions were consistent with the true intentions and recorded the times of errors. For a non-yielding vehicle, the time of collisions was counted as a supplement. 

The error rate for an eHMI design was calculated as the number of times that the participants misunderstood the intention of AV with the eHMI design, divided by the total examined times (each design was examined by each participant three times, the total number $N=31\times3=93$), as aforementioned in Section \ref{Sec:Depenedent_variables}.

Because “Error” is a categorical variable (error = 1, no error = 0), we used the chi-square test to evaluate whether there is an association between different eHMI designs and “Error”.

Given that each participant encountered AVs with each eHMI design three times, perceptions and behaviors of participants may vary with each time. To investigate whether there is an interaction between the different eHMI designs and the times of encounter, we conducted a two-way repeated measures ANOVA for analysis. We tested the data for normality of the residuals and sphericity assumptions with Shapiro-Wilk’s and Mauchly’s tests, respectively. If the assumption of sphericity was violated, we corrected the tests using the Greenhouse-Geisser method. For significance testing, we used two-way repeated-measures ANOVA to identify significant effects and applied Bonferroni-corrected t-tests for post-hoc analysis. Further, we report the generalized ETA squared $\eta_G^2$ as a measure of the effect. If the homogeneity of variance assumption was violated, Welch's  ANOVAs were used. If normality was violated according to the results of Shapiro-Wilk's tests, we applied the Aligned Rank Transform (ART) as proposed by Wobbrock et al. \cite{elkin_aligned_2021}.

Further, we plotted the average data as line graphs, as shown in Fig. \ref{fig:Y-Learn} and Fig. \ref{fig:N-Learn}. In a line graph, the X-axis represents the 1st, 2nd, and 3rd time, and the Y-axis represents the average of the 31 participants' performance.

Additionally, we applied the Student's t-Tests for physiological data to examine the difference between the baseline and each level. If normality was violated, Wilcoxon Rank-Sum tests were used.

\subsection{Results}

\subsubsection{Yielding Vehicles}
\
\newline
\indent \textbf{Error Rate: }The error rates are summarized in Table.\ref{tab:Y-ErrorRate}
There is a significant association between different eHMI designs and “Error” (${\chi}^2$(5) = 31.445, p < .001). The intentions of AVs without any HMI design were inaccurately understood 16 times. Y2, Y3, and Y4 were almost not misinterpreted by the participants (N=2, 1, 1). The frequency of misinterpretation of the intentions of Y1 and Y0 is relatively low (N=7, 7). 

\begin{table}[h]
    \caption{Error Rate for Yielding eHMI}
      \label{tab:Y-ErrorRate}
    \centering
    \begin{tabular}{ccccccc}
    \toprule
        Yielding eHMI design & Y1 & Y2 & Y3 & Y4 & Y0 & Y-Base \\ 
        \midrule
        Error Rate & 7/93 & 2/93 & 1/93 & 1/93 & 7/93 & 16/93 \\ 
        \bottomrule
    \end{tabular}
\end{table}

\textbf{Difficulty in Comprehension:} A significant main effect was observed ($F(5, 150)= 22.098, p < 0.001, \eta_G^2=0.250$) for eHMI designs on the difficulty in comprehension for yielding vehicles.  There was no significant main effect for the times of encounter. 

But the critical finding was the significant interaction effect for eHMI designs and encounter times ($F(10, 300)= 2.182, p = 0.041, \eta_G^2=0.019$). For Y0, there is a significant difference between the first time and the third time ($p = 0.002$), with a lower difficulty in comprehension for the third time.

\begin{figure}[h]
    \centering
    \subfigure[Difficulty in Comprehension]{
        \includegraphics[width=2.5in]{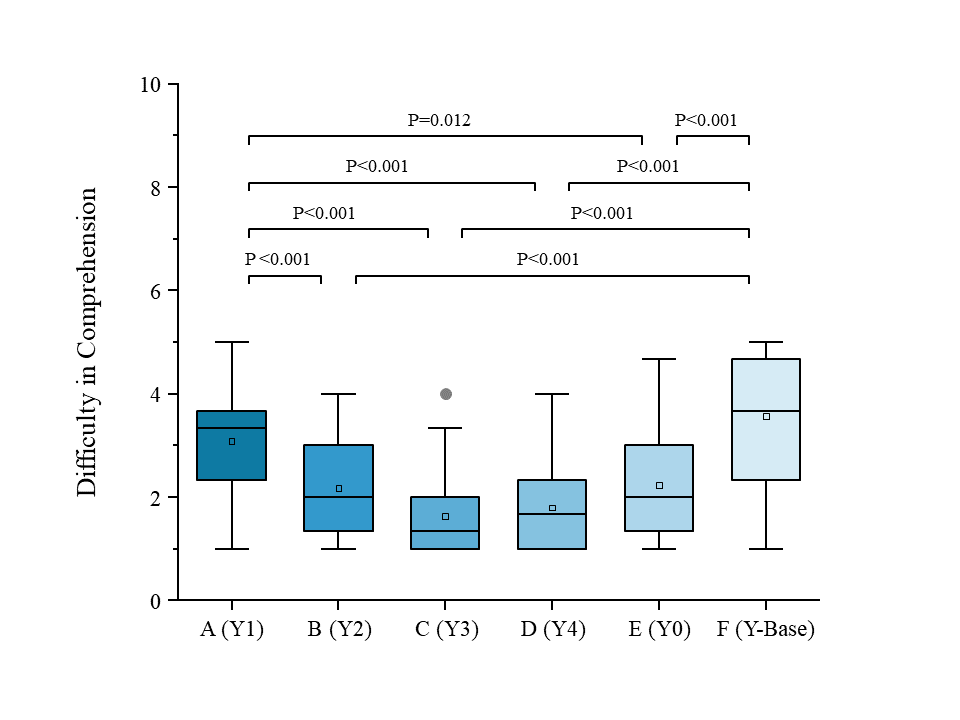}
        \label{fig:Y-Comprehension}
    }
    \subfigure[Perception of Danger]{
        \includegraphics[width=2.5in]{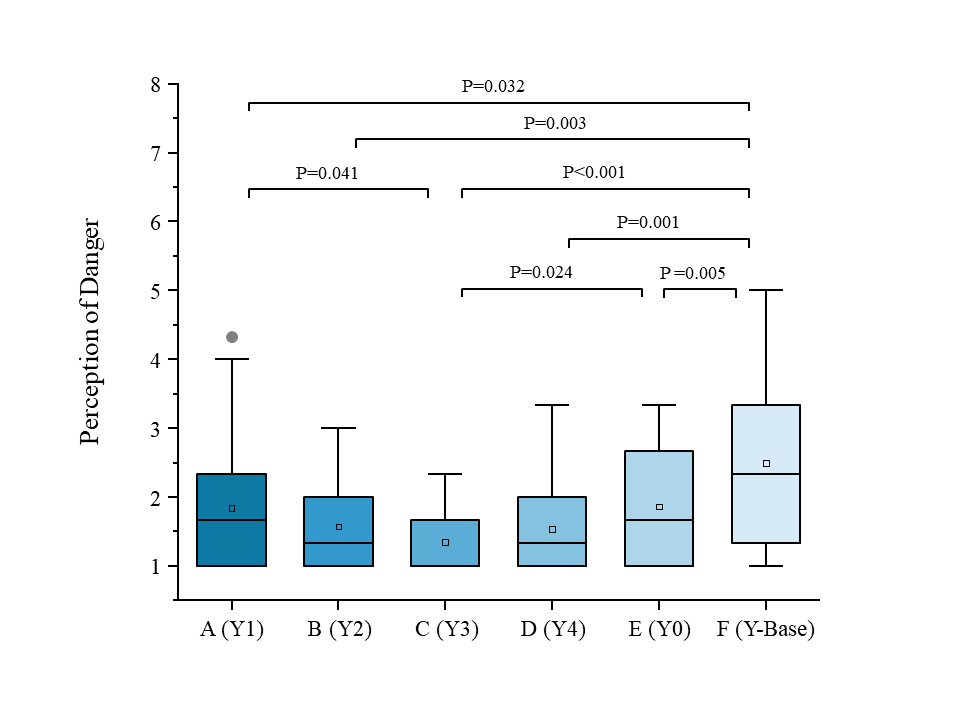}
        \label{fig:Y-Danger}
    }
    \quad    
    \subfigure[Duration of Observation]{
        \includegraphics[width=2.5in]{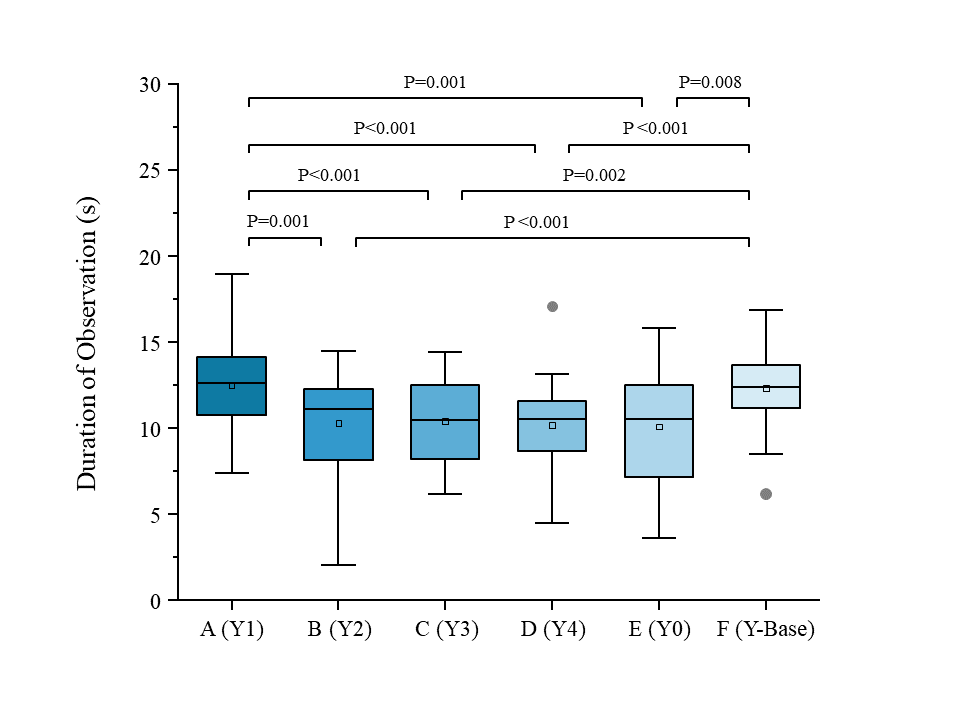}
        \label{fig:Y-ObservationTime}
    }
        \subfigure[Duration of Hesitation]{
        \includegraphics[width=2.5in]{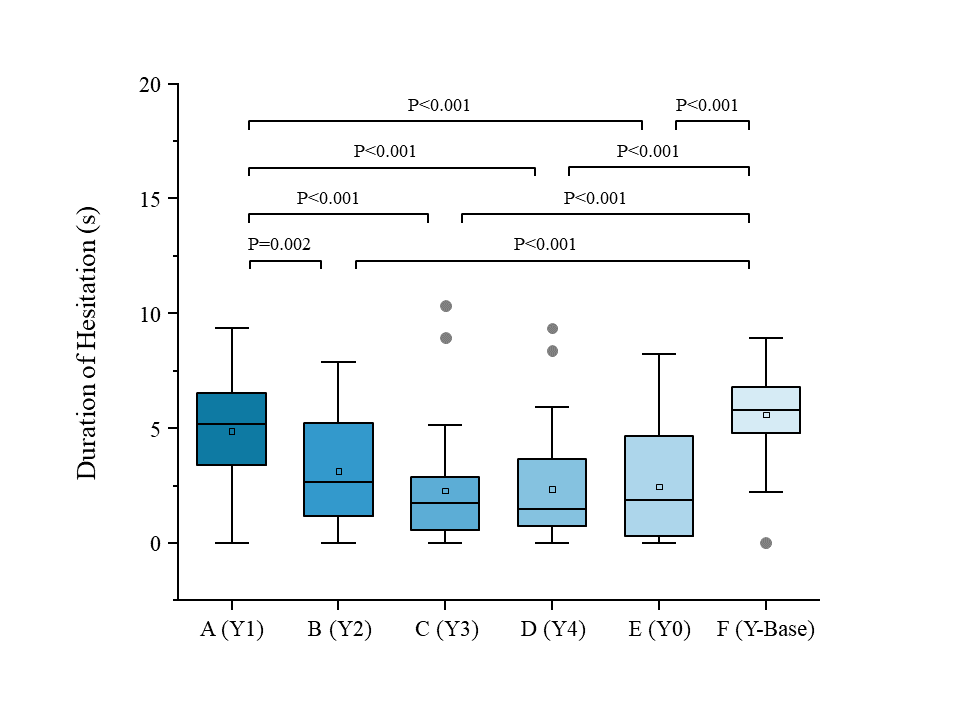}
        \label{fig:Y-HesitationTime}
    }
    \caption{Boxplots for (a),(b),(c),(d) by eHMI designs applied by AVs with yielding intentions}
    \label{fig:Y-Boxplot}
\end{figure}

Results of post-hoc tests indicate that the Y0, Y2, Y3, and Y4 are significantly easier to understand by participants than Y1 and Y-Base ($p < 0.05$). No significant difference is found between AV with Y1 and AV without any eHMI ($p > 0.05$). (Fig.\ref{fig:Y-Comprehension})

\begin{figure}[h]
    \centering
    \subfigure[Difficulty in Comprehension]{
        \includegraphics[width=2in]{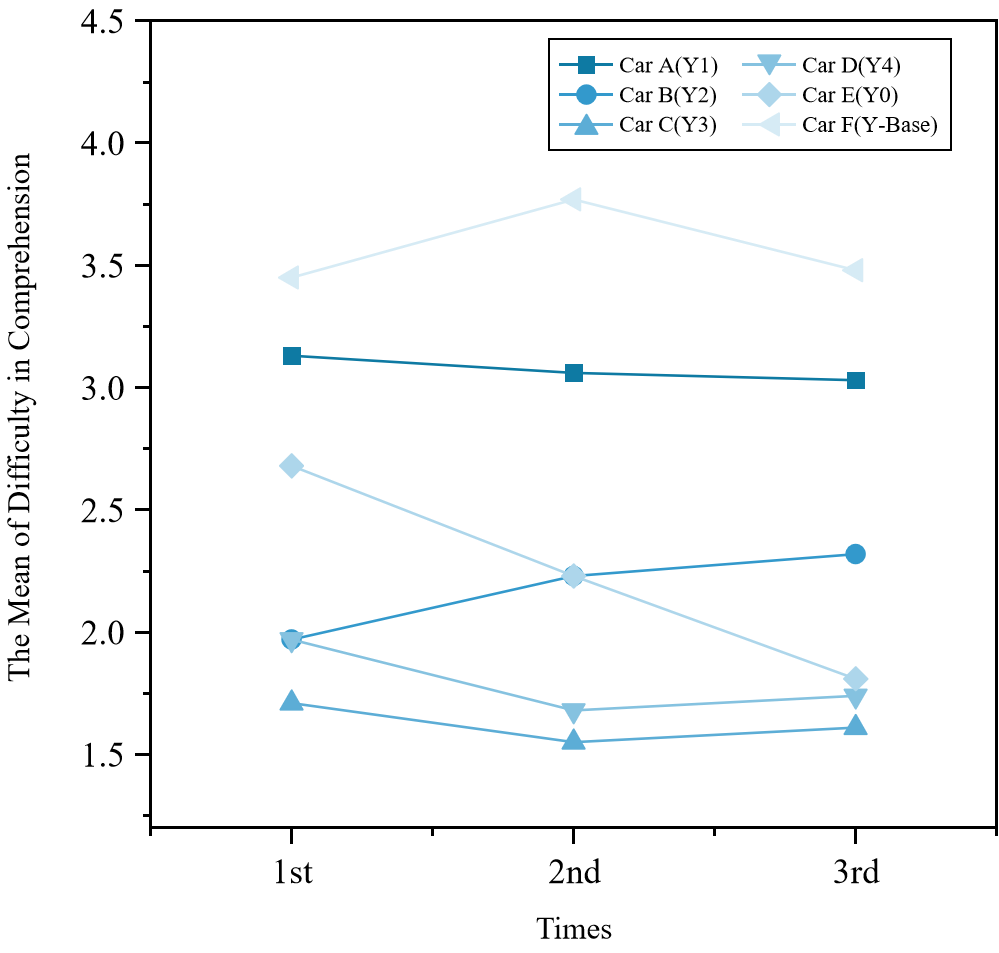}
        \label{fig:L-Y-Comprehension}
    }
    \subfigure[Perception of Danger]{
        \includegraphics[width=2in]{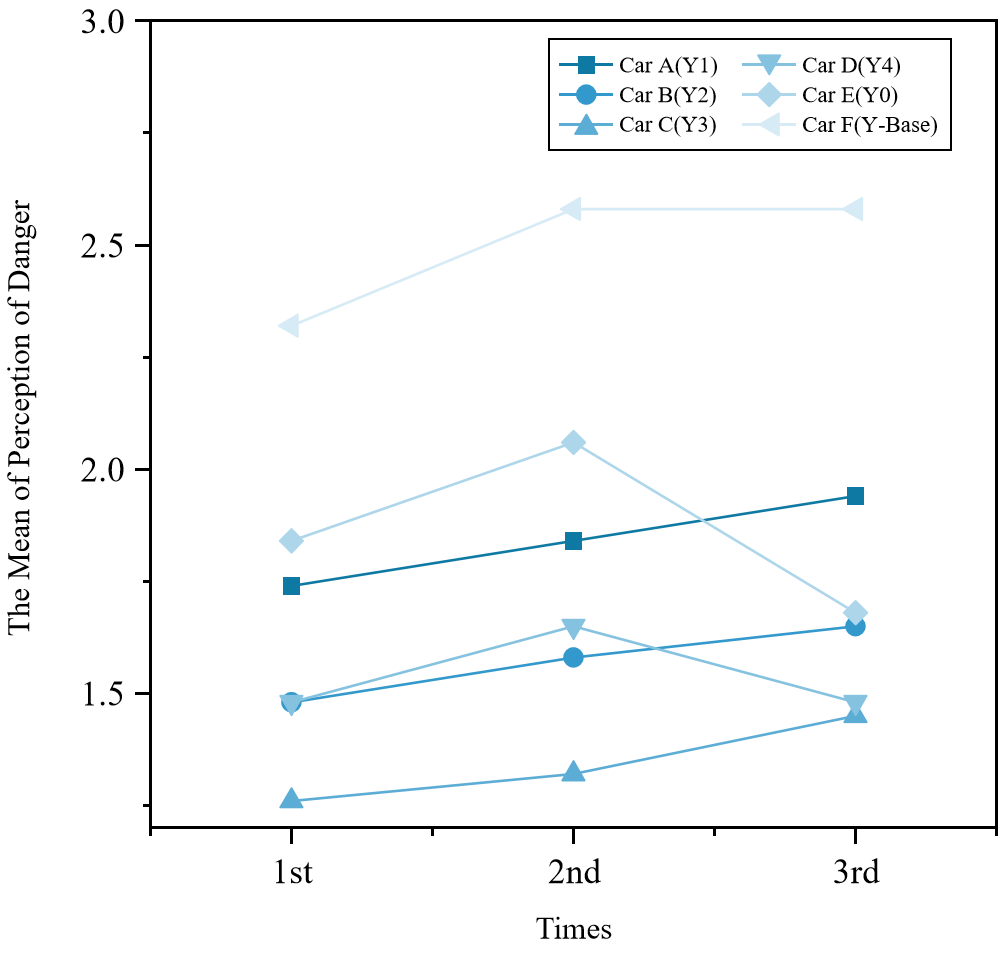}
        \label{fig:L-Y-Danger}
    }
   \quad    
    \subfigure[Duration of Observation]{
        \includegraphics[width=2in]{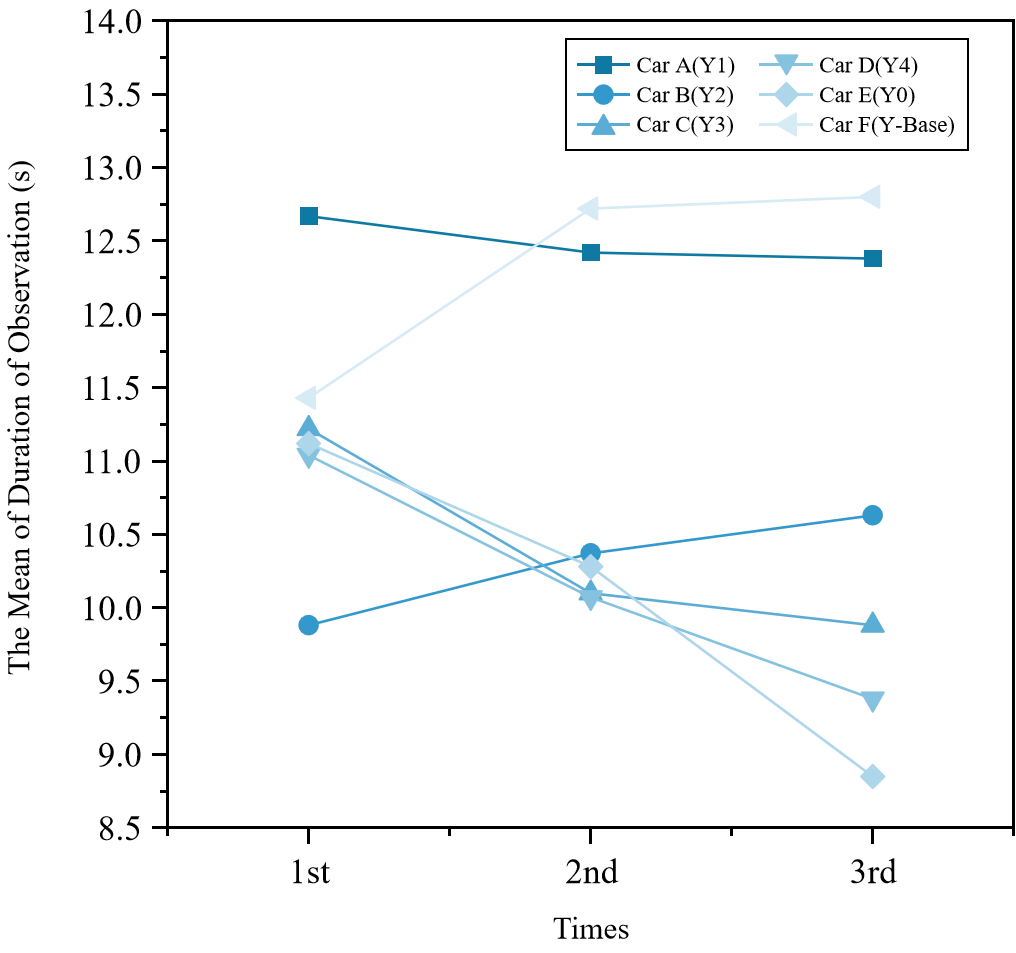}
        \label{fig:L-Y-ObservationTime}
    }
        \subfigure[Duration of Hesitation]{
        \includegraphics[width=2in]{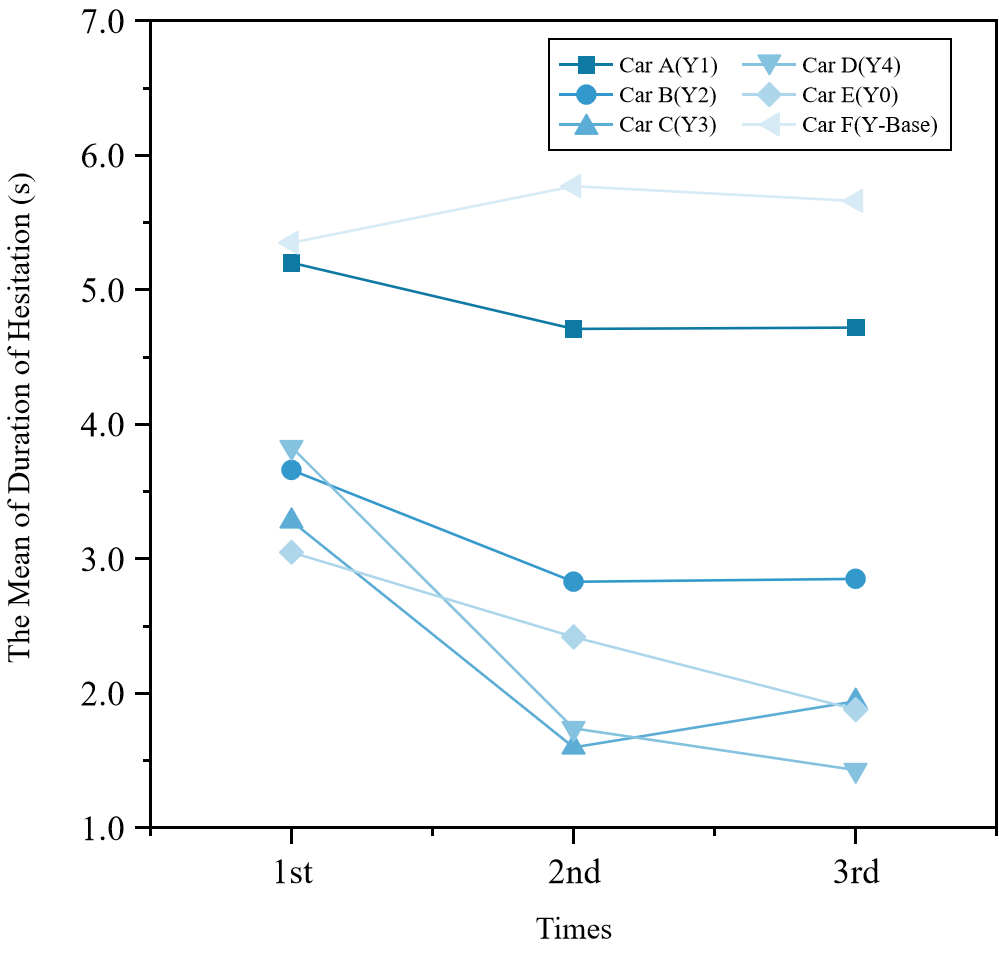}
        \label{fig:L-Y-HesitationTime}
    }
    \caption{Participants' performances about (a),(b),(c),(d) by encounters AVs with yielding intentions}
    \label{fig:Y-Learn}
\end{figure}

 Among yielding vehicles,  the average level of difficulty in perceiving the intentions of AVs without any eHMI design was found to be the highest and had slight variations among the three encounters.
The yielding intentions of AVs with Y1 are relatively hard to understand either. As the number of encounters increased, the decrease in comprehension difficulty was not noticeable. 
The difficulty of obtaining yielding intentions from AVs with Y3 is the lowest, followed by Y2, regardless of the number of encounters.
The level of comprehension difficulty for Y0 exhibits an approximately linear decrease, but for Y2 has an upward trend with increasing repetitions. (Fig.\ref{fig:L-Y-Comprehension})

\textbf{Perception of Danger:} There are significant differences ($F(5, 150)= 10.989, p <0.001, \eta_G^2=0.114$) between different designs on the perception of danger for yielding vehicles. There was no significant main effect for the times of encounter and no significant interaction effect for eHMI designs and encounter times. 

In contrast to AVs without eHMI applications, the applications of all five types of eHMI designs result in significant decreases in people's sense of danger compared with Y-base, based on the results of post-hoc tests ($p < 0.05$). (Fig.\ref{fig:Y-Danger})

It was shown that participants perceived AVs without an eHMI design as the most hazardous, while AVs equipped with Y3 elicited the highest sense of safety among participants, regardless of the number of encounters. (Fig.\ref{fig:L-Y-Danger})

\textbf{Duration of Observation:} A significant main effect was found ($F(5, 150)= 12.002, p < 0.001, \eta_G^2=0.092$) for eHMI designs on the duration of observation for yielding vehicles. There was no significant main effect for the times of encounter. 

We found a significant interaction effect for eHMI designs and encounter times on the duration of observation for yielding vehicles ($F(10, 300)= 2.900, p = 0.010, \eta_G^2=0.029$).
For Y4, there is a significant difference between the first time and the third time ($p = 0.003$), with a lower duration of observation for the third time. For Y0, there is a significant difference between the first time and the third time ($p = 0.027$), with less duration of observation for the third time.

In comparison to AV without eHMI design, the implementation of Y0, Y2, Y3, and Y4 demonstrates significant benefits in terms of decreasing the observation time ($p < 0.05$). However, the application of Y1 takes even longer than the time spent observing when compared to AV without eHMI design ($p > 0.05$). (Fig.\ref{fig:Y-ObservationTime})

Participants need to spend a greater amount of time to observe the intents of AVs equipped with Y1, compared to those without an eHMI design, during their initial encounter, and the duration of observation indicates a slight reduction as the number of repetitions increases. For AVs without an eHMI design, observing time demonstrates an upward trend with increasing repetitions. The duration required to figure out the intentions of AVs with Y2 is minimal during the initial encounter, but it gradually lengthens during subsequent encounters. The time required to observe Y3, Y4, and Y0 shows a noticeable decrease as the number of encounters rises. During the last meeting, it was found that the participants were able to comprehend the intentions of AVs equipped with Y0 in the shortest amount of time. (Fig.\ref{fig:L-Y-ObservationTime})

\textbf{Duration of Hesitation:} A significant main effect in the duration of hesitation time was observed ($F(5, 150)= 23.674, p < 0.001, \eta_G^2=0.202$). There was also a significant main effect for the times of encounter ($F(2, 60)= 12.560, p < 0.001, \eta_G^2=0.0.028$).

The critical finding was the significant interaction effect for eHMI designs and encounter times ($F(10, 300)= 3.352, p = 0.003, \eta_G^2=0.0.023$). For Y3 and Y4, there is a significant difference ($p < 0.05$), with less duration of hesitation for the second time and third time.

The utilization of Y0, Y2, Y3, and Y4 in comparison to AV without eHMI design proves to be significantly advantageous in terms of reducing the amount of time of hesitation ($p < 0.05$). Nevertheless, the application of Y1 fails to produce any significant distinction when compared to AV without eHMI design ($p > 0.05$). (Fig.\ref{fig:Y-HesitationTime})
Post-hoc analysis revealed significant differences in the duration of hesitation between the first and second encounters as well as between the first and third encounters ($p < 0.05$). It indicates that participants were able to respond more quickly when encountering an eHMI design that they had previously encountered.

For yielding vehicles, we found that pedestrians reacted worst to AVs without any eHMI design (Y-base). The hesitation time for the next two encounters rises compared to the performance for the first encounter. The performance of participants when confronted with Y1 is similar to that of Y-base, exhibiting a little improvement in the subsequent two encounters. However, it remains the worst among all eHMIs. The duration of hesitation indicates a discernable decreasing trend for Y2, Y3, Y4, and Y0. The pedestrians performed best in deciding their actions when they met AVs with Y0 in the first encounter. When considering the overall performance of the three encounters, it can be observed that Y4 shows the most pronounced Learning during experiencing. (Fig.\ref{fig:L-Y-HesitationTime})

\subsubsection{Non-Yielding Vehicles}
\
\newline
\indent \textbf{Error Rate: } The Error Rate and Collision Rate were summarized in Table.\ref{tab:N-ErrorRate}. 
There is a significant association between between different eHMI designs and “Error” (${\chi}^2$(5) = 28.049, p < 0.001). There is a significant association between different eHMI designs and “Collision” (${\chi}^2$(5) = 12.638, p = 0.027).
When answering question No. 1, some participants tend to report the observed behavior shown by the AVs after crossing the roads rather than their true thoughts before crossing. This may lead to a possible situation where the collision rate is greater than the error rate.

The intentions of AVs without any eHMI design were incorrectly interpreted 17 times, which was similar to the yielding AVs. The N4 eHMI design had the lowest number of misunderstandings, only eight times, compared to other eHMI designs. The frequency of misinterpretation of the intentions of N1, N2, and N0 is relatively low (N=12, 17, 11). Particularly, N3 was the worst among all eHMIs, and even more difficult to accurately understand than AV without any eHMI (N=29).
\begin{table}[h]
    \caption{Error Rate for Non-Yielding eHMI}
      \label{tab:N-ErrorRate}
    \centering
    \begin{tabular}{ccccccc}
    \toprule
        Non-yielding\\eHMI design & N1 & N2 & N3 & N4 & N0 & N-Base \\ 
        \midrule
        Error Rate & 11/93 & 12/93 & 23/93 & 8/93 & 11/93 & 17/93 \\ 
        \midrule
        Collision Rate & 12/93 & 17/93 & 29/93 & 7/93 & 7/93 & 13/93 \\ 
        \bottomrule
    \end{tabular}
\end{table}

\textbf{Difficulty in Comprehension:} A significant main effect was found ($F(5, 150)= 14.818, p < 0.001, \eta_G^2=0.169 $) between designs on the difficulty in comprehension for non-yielding vehicles. There was no significant main effect for the times of encounter and no significant interaction effect for eHMI designs and encounter times. 

\begin{figure}[h]
    \centering
    \subfigure[Difficulty in Comprehension]{
        \includegraphics[width=2.5in]{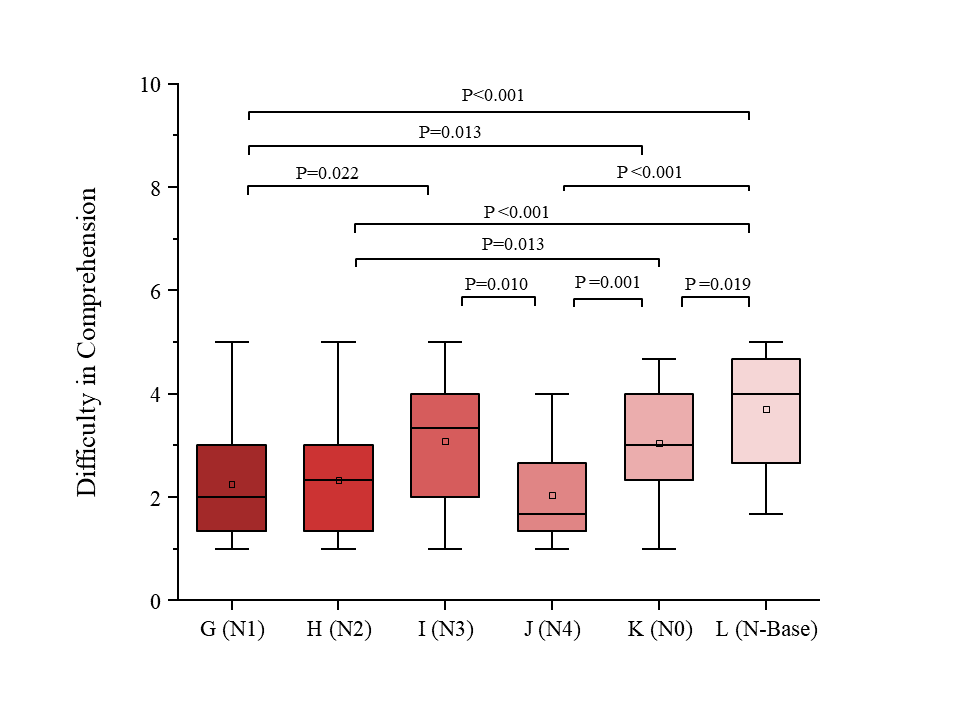}
        \label{fig:N-Comprehension}
    }
    \subfigure[Perception of Danger]{
        \includegraphics[width=2.5in]{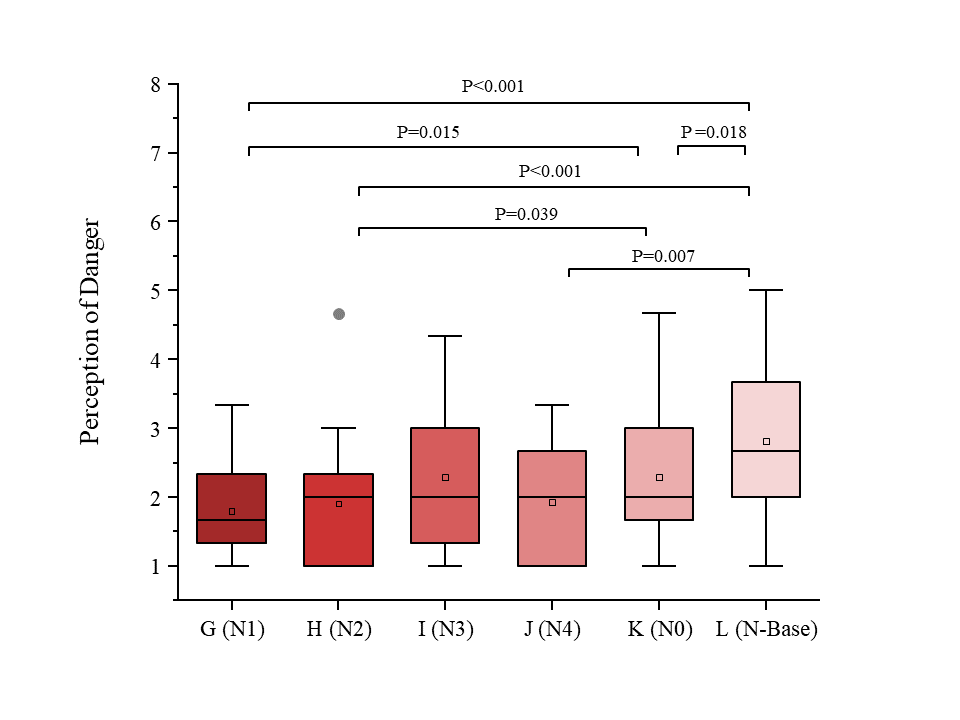}
        \label{fig:N-Danger}
    }
    \quad    
    \subfigure[Duration of Observation]{
        \includegraphics[width=2.5in]{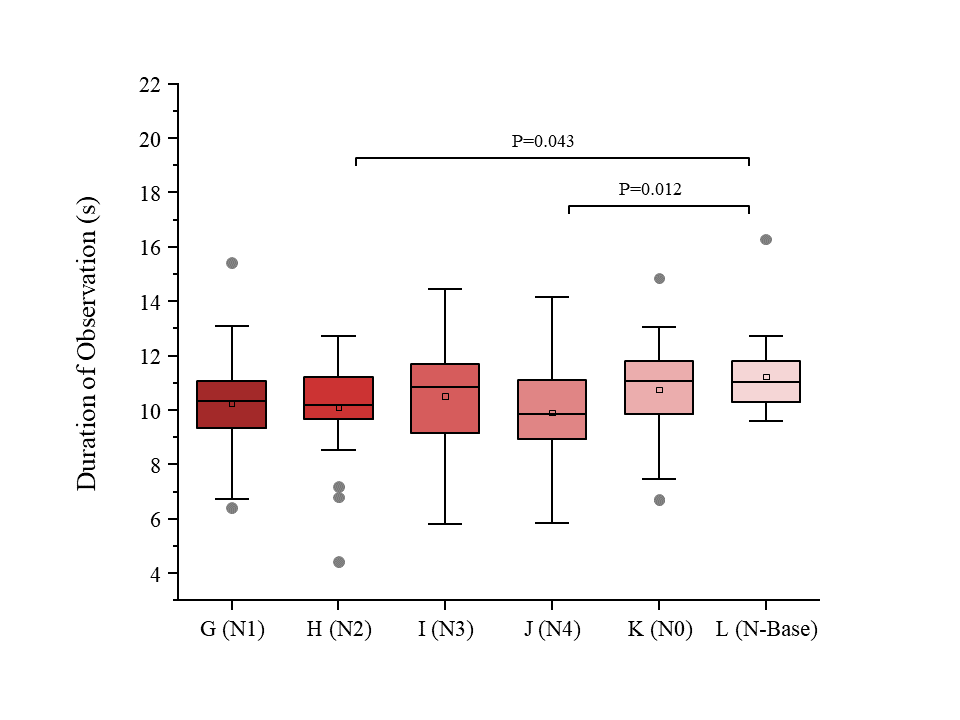}
        \label{fig:N-ObservationTime}
    }
        \subfigure[Duration of Hesitation]{
        \includegraphics[width=2.5in]{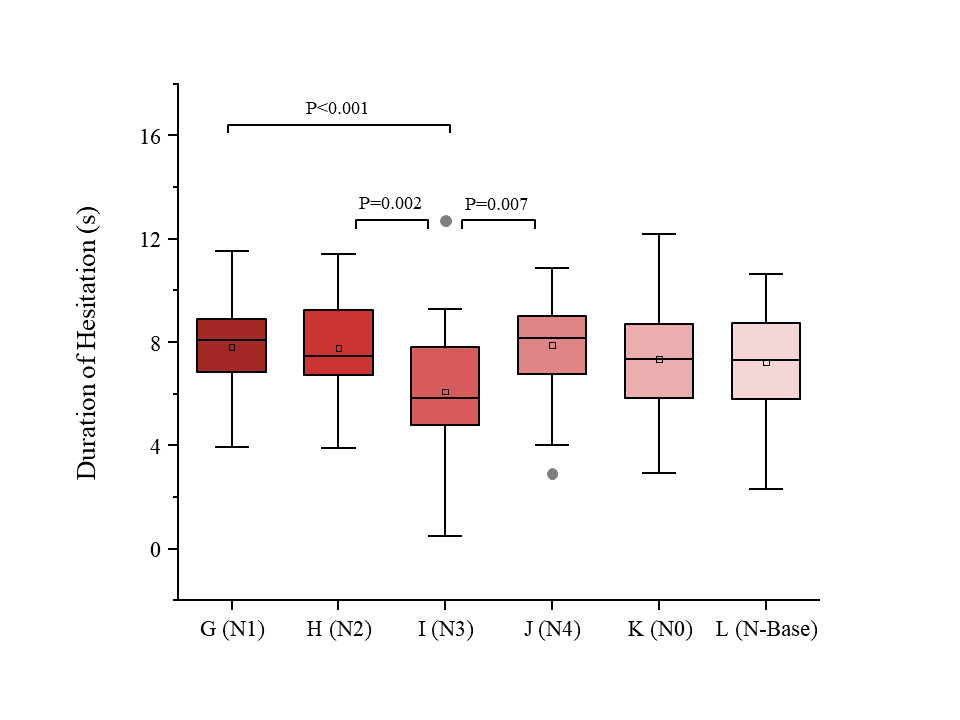}
        \label{fig:N-HesitationTime}
    }
    \caption{Boxplots for (a),(b),(c),(d) by eHMI designs applied by AVs with non-yielding intentions}
    \label{fig:N-Boxplot}
\end{figure}

Results of post-hoc tests indicate that the N0, N1, N2, and N4 are significantly easier to understand by participants than N-Base ( $p < 0.05$). As compared to AV without eHMI, N3 has no advantages in terms of cognitive comprehensibility ($p > 0.05$). (Fig.\ref{fig:N-Comprehension})

\begin{figure}[h]
    \centering
    \subfigure[Difficulty in Comprehension]{
        \includegraphics[width=2in]{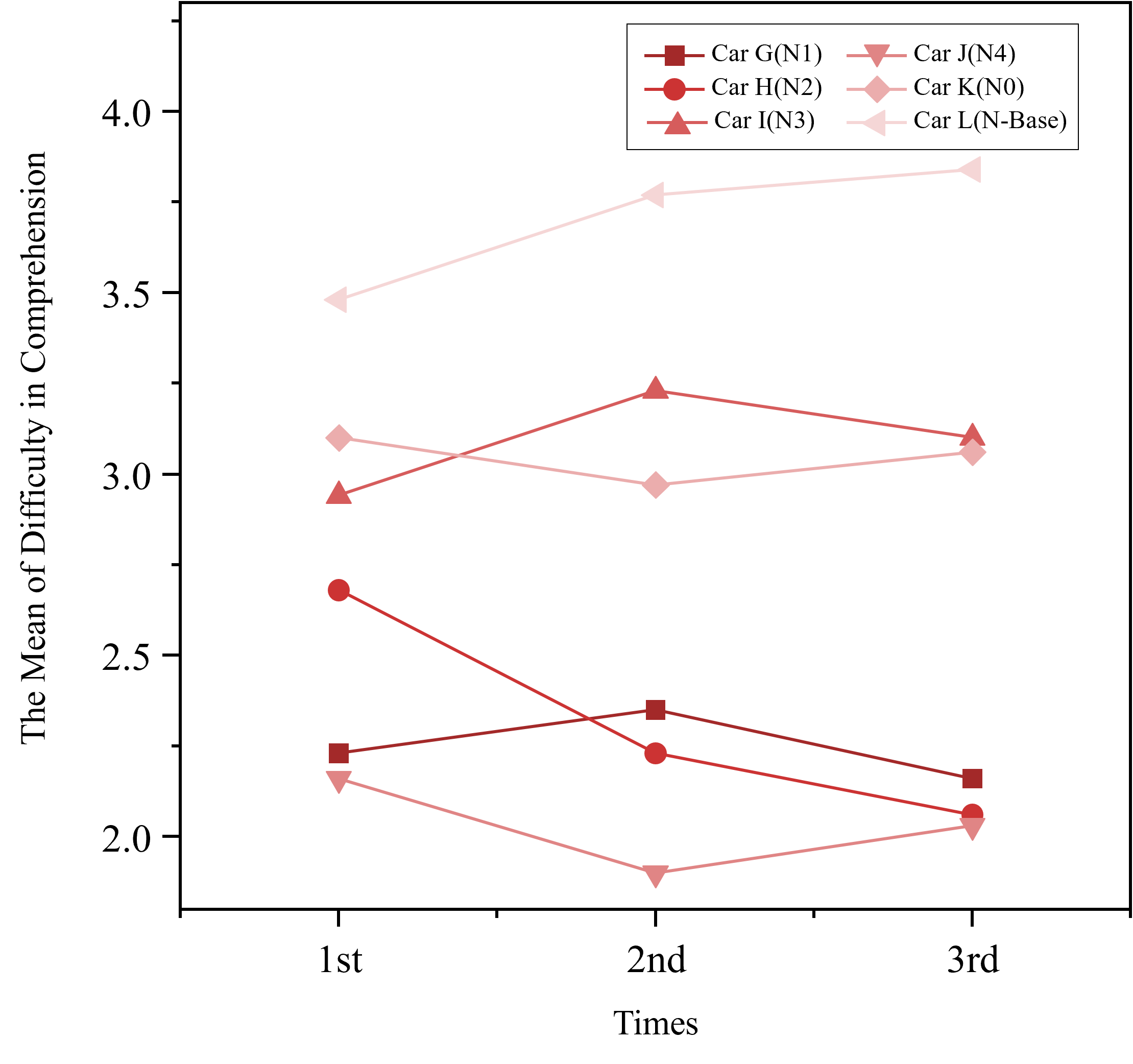}
        \label{fig:L-N-Comprehension}
    }
    \subfigure[Perception of Danger]{
        \includegraphics[width=2in]{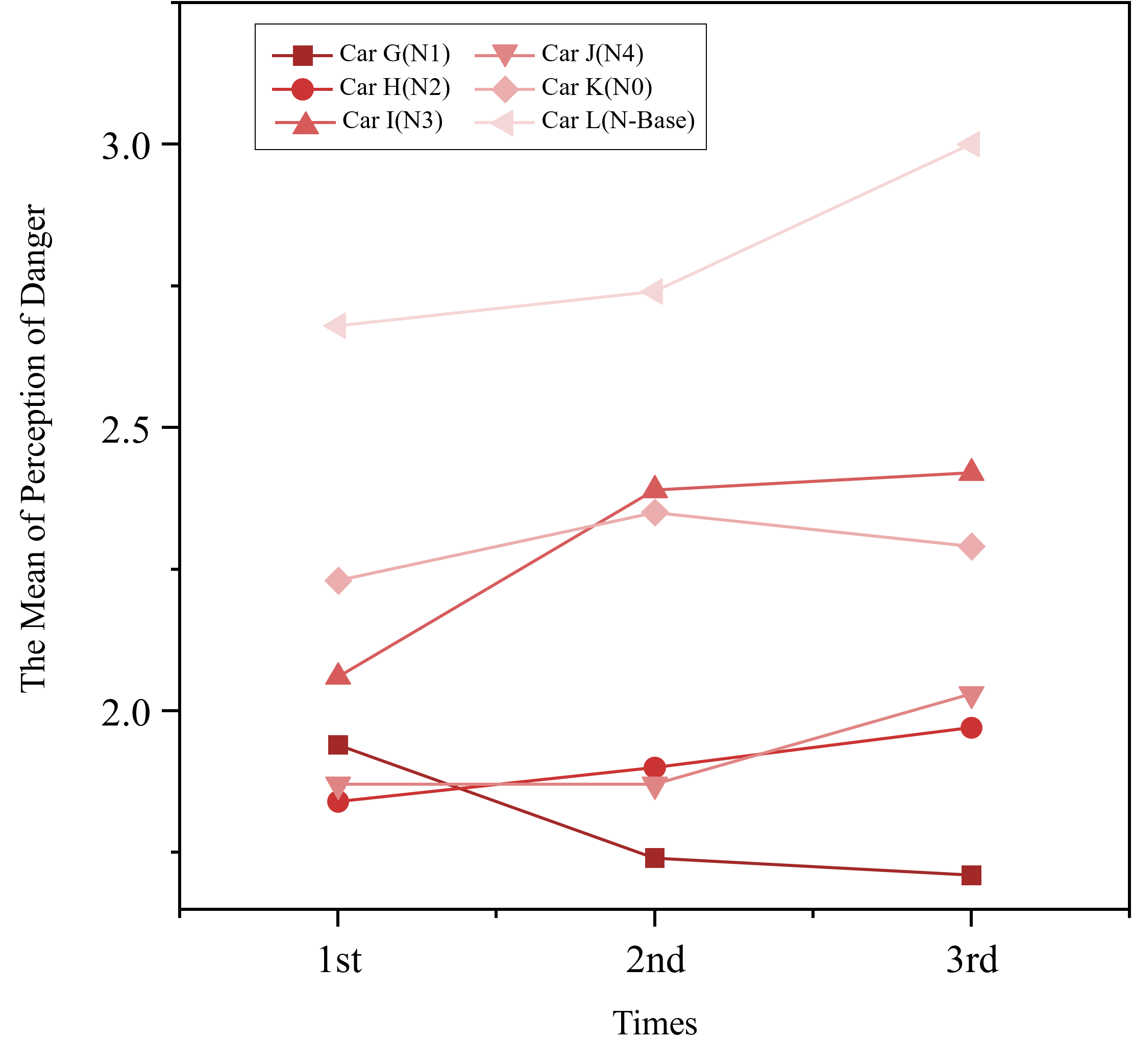}
        \label{fig:L-N-Danger}
    }
   \quad    
    \subfigure[Duration of Observation]{
        \includegraphics[width=2in]{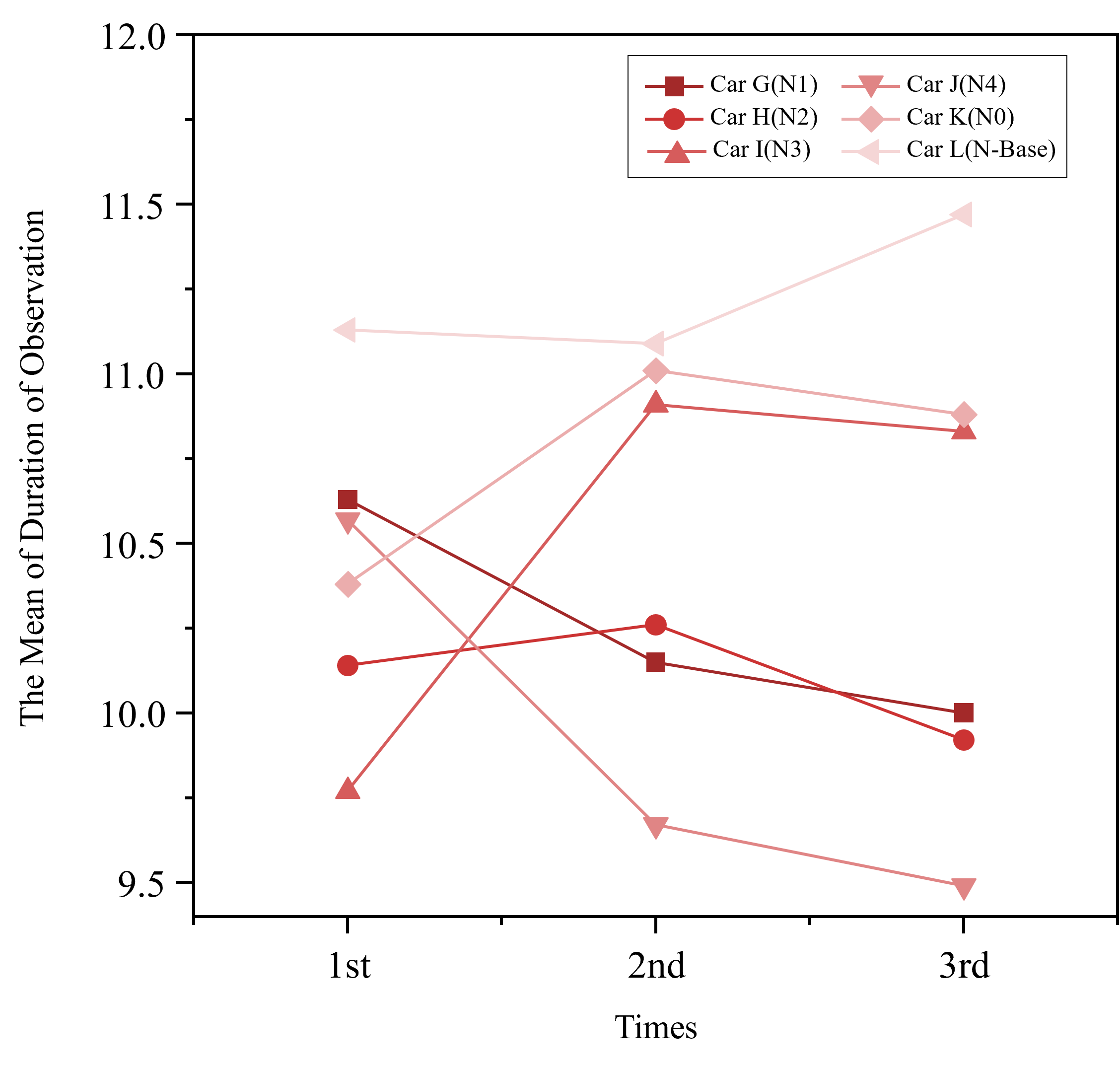}
        \label{fig:L-N-ObservationTime}
    }
        \subfigure[Duration of Hesitation]{
        \includegraphics[width=2in]{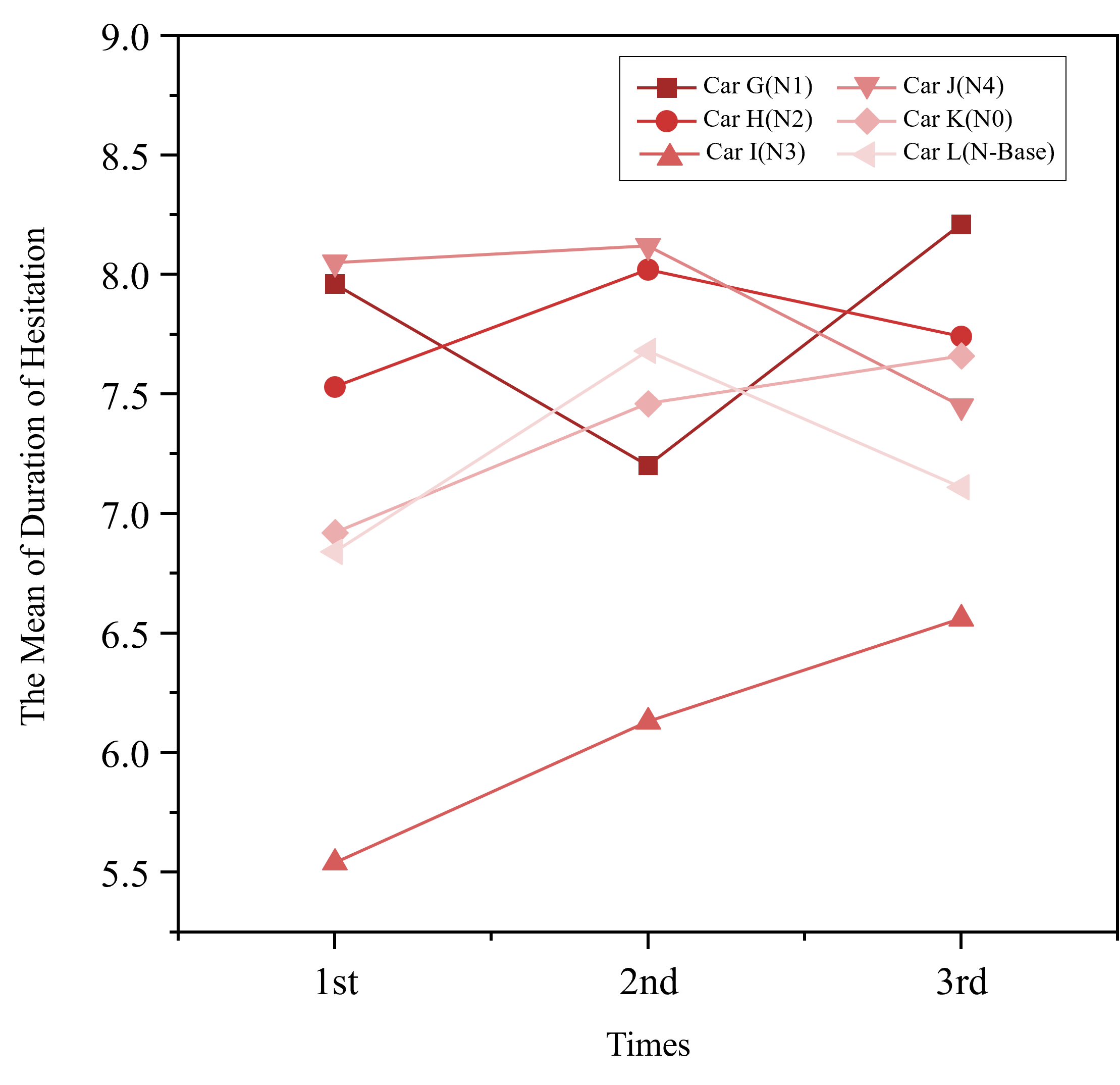}
        \label{fig:L-N-HesitationTime}
    }
    \caption{Participants' performances about (a),(b),(c),(d) by encounters AVs with non-yielding intentions}
    \label{fig:N-Learn}
\end{figure}

 Among non-yielding vehicles,  the average level of difficulty in perceiving the intentions of AVs without any eHMI design was found to be the highest and has an upward trend with increasing repetitions. AVs with N0 and N3 show similar levels of understanding difficulty.
The non-yielding intentions of AVs with N2, N1, or N4 can be relatively easy to perceive by participants among which, with N4 being the most easily understandable. The level of comprehension difficulty for N2 tends to diminish with increasing repetitions. (Fig.\ref{fig:L-N-Comprehension})
\textbf{Perception of Danger:} A significant main effect was found ($F(5, 150)= 9.703, p < 0.001, \eta_G^2=0.078$) between designs on the perception of danger for non-yielding vehicles. There was no significant main effect for the times of encounter and no significant interaction effect for eHMI designs and encounter times. 
The use of N0, N1, N2, and N4 on AVs results in significant improvements in participants' perception of safety when compared to AV without any eHMI design ($p < 0.05$). The application of N3 does not have a significant impact on the perceptions of risk ($p > 0.05$). (Fig\ref{fig:N-Danger})

It was shown that participants perceived AVs without an eHMI design as the most hazardous.  It has been observed that specifically for N1, the perception of danger of participants tends to decrease as the number of encounters increases. For the 2nd and 3rd encounters, N1 is regarded as the most secure eHMI design for pedestrians. Participants experience a similar sense of danger when they encounter AVs with N2 and N4. (Fig.\ref{fig:L-N-Danger})

\textbf{Duration of Observation:} A significant main effect was found ($F(5, 150)= 3.504, p = 0.011$) between designs on the duration of observation for non-yielding vehicles. There was no significant main effect for the times of encounter and no significant interaction effect for eHMI designs and encounter times. 

As compared to AV without eHMI design, the implementation of N2 and N4 is significantly helpful in shortening the observation time ($p < 0.05$). The utilization of other eHMI designs does not provide any significant advances ($p > 0.05$) towards reducing the time spent observing. (Fig.\ref{fig:N-ObservationTime})

 Among non-yielding vehicles,  the average duration of observing the AVs without any eHMI design was found to be consistently the longest, irrespective of the number of times it is encountered.  The time needed to understand the intentions of AVs equipped with N3 is shortest on the first encounter but increases when encountered again.  The duration required to observe AVs with N1 or N4 decreases with increasing repetitions. When met for the second or third time, N4 requires the shortest duration for comprehension. (Fig.\ref{fig:L-N-ObservationTime})

\textbf{Duration of Hesitation: } A significant main effect in the duration of hesitation time was observed  ($F(5, 150)= 8.000, p < 0.000, \eta_G^2=0.049$) between designs for non-yielding vehicles.  There was no significant main effect for the times of encounter and no significant interaction effect for eHMI designs and encounter times. 

The duration of hesitation when meeting with Avs equipped with N3 is noticeably lower compared to other counterparts. Nevertheless, it does not imply that N3 is the most effective. In contrast, pedestrians often misinterpret the non-yielding intentions of AVs equipped with N3, leading them to initiate crossing the crosswalk before AVs arrive. (Fig.\ref{fig:N-HesitationTime} and Fig.\ref{fig:L-N-HesitationTime})

\subsubsection{Physiological State}
\
\begin{figure}[h]
    \centering
    \includegraphics[width=\linewidth]{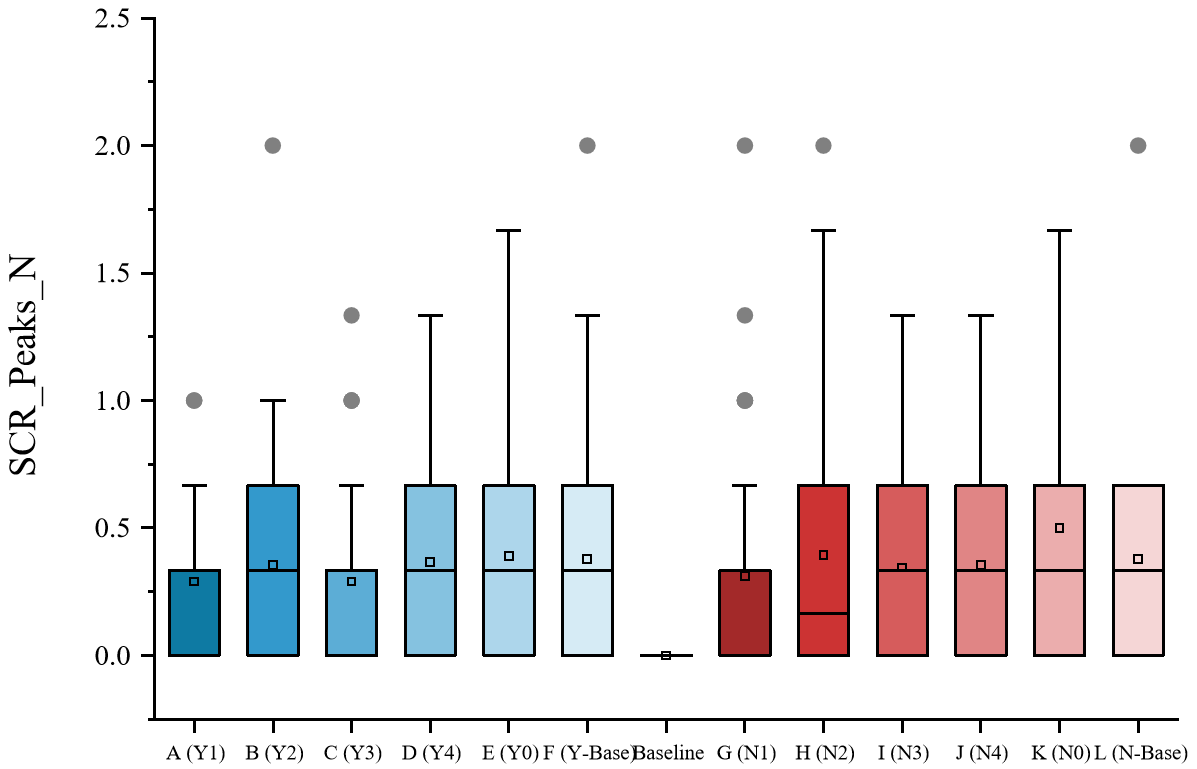}
    \caption{SCR\_Peaks\_N by eHMI designs}
    \label{fig:SCR}
\end{figure}

\indent \textbf{SCR\_Peaks\_N} (the number of SCR peaks) can indicate the number of times the participant was stimulated over a period of time~\cite{braithwaite2013guide}. The participants were not exposed to any external stimuli during the collection of their baseline physiological data. During the experiments, participants were stimulated in each trial. ANOVA reveals no statistically significant differences between designs.  For all eHMI designs, the SCR\_Peaks\_N is significantly different from the baseline ($p < 0.05$). 
(Figure\ref{fig:SCR})


 \textbf{HR} and \textbf{HRV\_RMSSD}  have no significant differences, neither between designs nor between each design and baseline.

\subsection{Qualitative Data}
In addition to quantitative data, qualitative feedback also enables us to gain insights from the subjective reasoning of participants. We conducted a thematic analysis of the responses to semi-structured interviews, which required participants to recall the process of answering questions and explain the reasons for ranking eHMI. At the end, there will be several questions about trust in autonomous vehicles and gesture interaction. The insights generated from this are discussed here, and some relevant participant quotes are provided. When recalling the situation of their experiment, participants often use various words to indicate gestures and lights, such as "driver", "red light", "green light", etc., to refer to the eHMI of the vehicle they saw and even imitate the gestures in memory during the interview to represent the eHMI they saw at the time. For the convenience of discussion, the numbering of gestures will be used in subsequent quotations to clarify the gestures mentioned by the participants, such as Y1, N1, etc. All quotes used in the paper were translated into English by the lead author and checked by other co-authors.
\\
\textbf{Recipient and content: "It must be gesturing towards me"}
\\
The key to whether pedestrians can effectively interact with autonomous vehicles is whether pedestrians can receive and understand the information conveyed by the vehicle eHMI. Before receiving information, pedestrians need to know that they are the Receiver of eHMI information. However, some participants mentioned that they did not believe that the car was sending messages to themselves during the experiment. As \textit{P27} said:
\begin{itemize}
    \item [\emph{(P27)}]\emph{"I'm not sure if the lights are giving people instructions or indicating the status of the car, or if they are warning other cars that it is about to brake. This confuses me because the results of these two different ways of understanding are opposite."}
    \item [\emph{(P15)}]\emph{“The lights are a little blurry; I'm not sure if they're for decoration or to express the meaning.”}
\end{itemize}

In addition to lights, some gestures have the potential to confuse people.
\begin{itemize}
\item[\emph{(P28)}:]\emph{ "At the crossroads, I wasn't looking straight at the front window of the car; instead, I was looking at it from an angle. Anyway, I don't think the palm of the hand like N2 is facing me."}
\item[\emph{(P32)}] \emph{"I don't think that Y1 is a gesture made for me."}
\end{itemize}

Nonetheless, the majority of participants stated that when they saw the gesture, they immediately recognized that the car or virtual "driver" was interacting with pedestrians.
\begin{itemize}
    \item [\emph{(P8)}]\emph{"I know it must be gesturing towards me, but the lights may not be."}
    \item [\emph{(P29)}]\emph{"I feel like I'm talking to a real driver when I use gestures, but I feel like I'm in the future when I use lights, which look like road signs."}
\end{itemize}

On the other hand, the content of the information conveyed by eHMI is also very important.
The utilization of lights, specifically red and green, for conveying information is characterized by its simplicity, as it is limited to the binary options of "yes" or "no", "go" or "stop". In contrast, gestures possess the capacity to convey a greater amount of complicated information due to their intricate series of motions, encompassing factors such as the direction of gesture movement, frequency of repetition, etc.
\begin{itemize}
    \item [\emph{(P8)}]\emph{"When I see a hand gesture, I always think that the hand is an avatar of myself, so if the direction of movement of the gesture is consistent with the direction of my walking, it will be easy to understand."}
    \item[\emph{(P32)}]\emph{ "The best gesture is to move from the outside to the inside, from the pedestrian's position to the driver's position, which can indicate the meaning of the pedestrian passing in front of the driver."}
\end{itemize}

Nevertheless, the complexity may not always be advantageous for all types of information required to be conveyed, as \emph{P17} told us:
\begin{itemize}
    \item [\emph{(P17)}]\emph{"In my opinion, the gesture means non-yielding is stopping a pedestrian, so the gesture shouldn't be a moving action, it should be unmoving."}
\end{itemize}

When indicating the intention not to yield, several participants agreed on the use of a static gesture rather than a dynamic one on an eHMI system. They felt that a dynamic gesture could result in confusion, especially in the case of N3, which shows a waving motion closely resembling that of Y2.
\\
\textbf{Interaction distance and the observability of eHMI: "It was just too little to notice from a distance"}
\\
Through conducting experiments and interviews, it was observed that pedestrians consider "distance" to be an essential factor, which is consistent with the conclusion of Dey et al. \cite{dey2020distance}. The identification of a vehicle's intent at an early stage allows pedestrians to cross the street at a safer distance from the vehicle, resulting in increased safety and time efficiency. Nevertheless, it is hard to achieve in both virtual reality (VR) tests and real-world scenarios——the human eye has visual limitations in seeing objects that are far away clearly. When driving close, their specific intentions become evident within a few seconds, either by coming to a stop or passing by. Some of the participants still judge the vehicle's intention by observing its movement behavior, such as waiting for the vehicle to either pass by or come to a stop before proceeding to cross the street. They said:
\begin{itemize}
    \item [\emph{(P7)}]\emph{“The distance is too far to see clearly. Only when the car is very close to me can I see clearly. But it's already too late. The car is about to stop.”}
    \item[\emph{(P12)}] \emph{"Although you told me that whether a car yields or not, their speed is the same, I am still used to judging based on the speed of the vehicle."}
\end{itemize}

Avoiding this quandary is also challenging for gesture-based eHMI. Gestures without a sufficient range of motion exhibit visual similarity when observed from a distance. In cases where these gestures have distinct intentions, their possible consequences can be quite severe.
\begin{itemize}
    \item [\emph{(P16)}]\emph{"Y1 and N2 are difficult to distinguish at a long distance, I can only see one hand.."}
    \item [\emph{(P22)}]\emph{"Y2 and N3 look exactly the same from a distance. All you can see is a person waving, not knowing whether they're trying to yield or not."}
    \item [\emph{(P2)}]\emph{"I can clearly see gestures with a wide range of motion, such as N4 and Y3, but not small movements like Y4 and Y1."}
    \item [\emph{(P24)}]\emph{"I had a driver's license, so I recognized the N2 when I first saw it closely; I realized it was a traffic cop's hand signal, but it was just too little to notice from a distance."}
\end{itemize}

It can be found that an extended visual distance enhances pedestrians' ability to comprehend the intentions of a vehicle, regardless of whether it involves a gesture or any other form of eHMI. Hence, for gesture-based eHMI, the significance of a wider range of motion, increased surface area of the arm exposed, and an adequately discernible interface can not be ignored.
\\
\textbf{Unfamiliar traffic signal: Green indicates “Go”~? Red indicates “Stop”~?}
\\
Although traffic regulations in China have been regulated and people have become accustomed to the fact that green indicates 'go' and red indicates 'stop', participants' performance on red and green lights was not at the expected levels, and participants' ratings of red and green lights varied widely.
Initially, the majority of the people involved in the experiment would guess as to whether the color of the eHMI light strip on the car corresponds with the color of the usual traffic lights, specifically green indicating 'go' and red indicating 'stop'.But they still need to try empirically before they can confirm the facts. It is consistent with the conclusion of Dey et al.~\cite{dey_color_2020}.
\begin{itemize}
    \item [\emph{(P32)}]\emph{"The red or green lights on the car cannot represent the intention of the vehicle, and the flashing lights give a sense of relaxation, regardless of the color."}
\end{itemize}

Participants who took a conservative approach to crossing the street found it easier to understand that the red light signals 'Stop' rather than the intention represented by the green light: 
\begin{itemize}
    \item [\emph{(P25)}]\emph{"A red light is better understood than a green light. If I see a red light, I will definitely not move, but even if it is a green light, I may not dare to walk because it is unknown to me, and I will feel afraid."}
    \item[\emph{(P26)}]\emph{"I may not understand the meaning of a green light, but I know that a red light means non-yield."} 
\end{itemize}

Furthermore, we discovered that in addition to their inconsistent meanings, red and green lights appeared to many participants to be very different in other ways. The participants had different levels of difficulty comprehending the red and green lights:
\begin{itemize}
    \item [\emph{(P18)}]\emph{"I'm not sure what the red light means; I'm not sure if it's a brake light; it could use a little more graphics."
    \item[\emph{(P20)}]\emph{"Red lights are easy to understand but can not be seen clearly. If a blue/green light suddenly appears, I can't confirm the intention of the vehicle."}} 
\end{itemize}

\textbf{Comprehensibility of gestures:"I immediately understood its meaning"}
\\
Participants explaining the reasons for the ordering in the interviews often mentioned that they often encountered a certain gesture in their own lives, which could either make the gesture simple to understand or cause confusion due to the common intent of the gesture in their lives. For example, Y1 can also indicate you're awesome, while N3 can mean hello, farewell, and so on. Participants said:
\begin{itemize}
    \item [\emph{(P27)}]\emph{"Y4 looks like a person walking. It's easy to understand. Y1 seems to be praising. I completely cannot understand its meaning." }
    \item [\emph{(P18)}]\emph{"On my first encounter with N3, I mistook the vehicle for an attempt to yield to my intention, and I crossed the road immediately, not expecting to be hit at all."}
    \item [\emph{(P5)}]\emph{"Y4 impressed me, but it was silly and not natural like it was for a child; it took more effort and was more complicated to understand."}
    \item [\emph{(P4}]\emph{"The most commonly used gesture by drivers is Y2, and I also use Y2 when driving."}
\end{itemize}

Although some of the gestures were not familiar to the participants, they were easy to learn and effective:
\begin{itemize}
    \item [\emph{(P9)}]\emph{"Although I have never seen a driver gesture to me like this (Y3 and Y4) in my life, I immediately understood its meaning when I first saw it today. I quickly learned the rest after trying the intersection once."}
    \item [\emph{(P28)}]\emph{"Gestures have a low learning cost and were well understood in today's experiment, whereas lights have a high learning cost and were not so well understood." }
\end{itemize}

Participants largely felt that the ineffective gestures well enough were the lowest-ranked of all eHMIs. However, when compared to an AV without any eHMI designs, the attitudes of participants varied.
\begin{itemize}
    \item [\emph{(P6)}]\emph{"I would find it dangerous if there is a conflict between the vehicle's speed information and the information displayed on the vehicle's eHMI."}
    \item[\emph{(P22)}] \emph{"Although gestures may be misunderstood, I think it is safer to have gestures, at least giving me the possibility to understand the intentions of the autonomous vehicle."}    
\end{itemize}

Although some gestures are ineffective at helping pedestrians understand the intent of the vehicle, these gestures will increase walkers' sense of security when crossing the street.
\\
\textbf{Communicating in a more friendly manner: "I am communicating with the vehicles"}
\\
Information conveyed through gestures is rich. In addition to the original meanings and directions, gestures also contain the driver's feelings and attitudes in daily life. In the experiment, although only the gesture animation was displayed on the external interface of the autonomous vehicle, the participants could still feel the "driver" emotion and think that they are communicating with a "person". According to the participants:
\begin{itemize}
    \item [\emph{(P31)}]\emph{ "In addition to their original meanings, gestures can also express the driver's mood and personality. I care about the driver's attitude when it comes to yielding, but I don't care about the driver's attitude when it comes to not yielding. Y1 and Y3 feel friendly, while Y2 waves back and forth to give a sense of urging, which I don't really like. The lights just have color differences and make it feel emotionless. Someone gesturing to me gives me a strong sense of security, even if I completely cannot understand the meaning of the gesture."}
    \item[\emph{(P26)}]\emph{"Some gestures always give me a feeling of being forced, such as Y2, which makes me uncomfortable. I want it to be my free choice whether I cross the road or not."} 
    \item [\emph{(P18)}]\emph{"Gestures enhance my sense of familiarity and give me psychological comfort because I feel that I am communicating with the vehicles."}
    \item[\emph{(P25)}]\emph{"Today was the first time I'd seen an AV with gestures; it wasn't as scary as I thought it would be, and I felt more relieved as if a driver is driving."} 
\end{itemize}

In the last questions of the interview, a majority (N=19) of the respondents indicated an increased level of trust in the autonomous vehicle after the experiment. Additionally, they expressed confidence in the vehicle's ability to effectively convey its intentions to pedestrians.
The majority of participants (N=26) showed agreement with the potential use of gestures as a means of facilitating communication between autonomous vehicles and pedestrians in future scenarios. This tendency was primarily influenced by good experiences during the experiments.

To understand the influencing factors of participants' ranking of gestures, We adopted grounded theory in the qualitative analysis\cite{GroundedTheory}, reviewing and labeling the emerging codes in an iterative process. The lead author conducted inductive coding of the data and discussed the resulting themes with other co-authors to identify principal themes from the data. The primary codes included: Clarity of gestures, familiarity with gestures, politeness of gestures, and so on. The whole research team reached a consensus on the final themes. 

\section{STUDY 2: Survey}
"Given the inherent costs of real-world validation testing, efficient methods for early stage concept assessment are highly desirable for narrowing in on designs that are promising and setting aside those less likely to prove out."\cite{fridman2017walk}The previous virtual reality experiments convincingly demonstrated the feasibility of utilizing gesture interaction in AVs and pedestrian interaction scenarios. To receive a broader range of viewpoints from the population and to gauge the acceptance of gesture interaction in a wider context, we conducted additional research utilizing a web-based questionnaire, building upon previous experiments.

\subsection{Survey Design}
This questionnaire research categorizes gestures into three dimensions for evaluation: clarity, familiarity, and politeness. Clarity refers to how clearly the gesture expresses intentions. Familiarity refers to how familiar the gesture is in daily life, and politeness refers to how polite the gesture is in expressing intentions. The Gesture Research Questionnaire collected basic demographic information of participants alongside evaluations of three gesture dimensions in a two-by-two comparison format. Participants were also asked to rate various gestures on the three dimensions (clarity, familiarity, politeness) using a two-by-two comparison rating. Technical term abbreviations were explained upon first use. The text adheres to all given principles and requires no further revision.

The Analytic Hierarchy Process (AHP) is a renowned method for decision-making, initially proposed by American operations researcher Saaty in the early 1970s. It involves decomposing decision-making elements into levels such as objectives, criteria, and options, which are then qualitatively and quantitatively analyzed through pairwise comparisons~\cite{forman_analytic_2001,saaty_exposition_1990,vaidya_analytic_2006}.
In this questionnaire study, an evaluation model was developed using the AHP method. Participants' scores for three-dimensional weights and multidimensional scores for each gesture were obtained by the two-by-two comparison method. To construct a unified judgment matrix, group opinions were harmonized by calculating the geometric mean. Additional analysis provided the three-dimensional scores for judging the gestures, as well as the scores of different gestures on the three dimensions. The results of the specific data analysis will be presented in the following sections. Additional information on questionnaire design and analysis can be found in the attached appendix\ref{appendix:Supplementary Note to Study 2}.

\subsection{Respondent Recruitment}
The survey has been approved by the local university’s Institutional Review Board (IRB). The survey was distributed via the Internet and was accessible to all respondents through an anonymous link. Since respondents played the role of pedestrians in the study, there was no screening for categories such as gender, age, and occupation, which allowed a wider range of people to participate in this survey. As an incentive to participate in the survey, all respondents whose questionnaires were verified as valid were rewarded with about \$0.7.

\begin{table}[h]
    \caption{Participants demographics of Study 2}
      \label{tab:Participants}
    \centering
    \begin{tabular}{ccc}
    \toprule
        Age group & Number &  Proportion \\ 
        \midrule
        12-18 & 26  & 7.37 \% \\
        19-25 & 206 & 58.36\% \\
        26-45 & 70 & 19.38\%\\
        46-60 & 45 & 12.75\%\\
        60 and above & 6 & 1.70\%\\
        \midrule
        Gender group & Number & Proportion\\
        \midrule
        Male & 164 & 46.46\%  \\ 
        Female & 189 & 53.54\%\\
        \bottomrule
    \end{tabular}
\end{table}

\subsection{Data Quality}
In this questionnaire study, several measures were taken to ensure data quality. Questions were phrased in everyday language to facilitate comprehension.  For better visual understanding, moving images with gestures were included in the questionnaire. Moreover, participants were informed that data collection was completely anonymous to prevent bias. Two questions were included in the questionnaire to eliminate the possibility of machine answers. Additionally, any questionnaires that were completed too quickly or took too long were deemed invalid (Less than 2 minutes or over 30 minutes) .

\subsection{Respondent Demographics}
We distributed 394 questionnaires to the general public via the Internet. We removed 41 invalid questionnaires, leaving 353 valid responses after a consistency test. The participants included 164 males and 189 females from various regions of mainland China. They ranged in age, with a larger proportion falling in the 18-25 age group. See the table for precise demographic information.

\subsection{Result}
After analyzing the data, we obtained the participants' ratings for each gesture. In the following, we will discuss the ratings of the different gestures and the differences between the three rating dimensions of the gestures in the different age groups.

\subsubsection{Total score for gestures}
\
\newline
For yielding gestures, the gesture with the highest score was Y2, followed by Y3, while Y1 and Y4 scored lower and closer, combining the weights of the three dimensions as well as the scores. In terms of clarity, gesture Y2 performed the best, followed by Y3, Y4, and Y1. The gesture with the best familiarity was Y2, followed by Y3, Y4, and Y1. Y4 scored highest on politeness, followed by Y1, while Y2 and Y4 scored lower and closer. Combining the above data, Y2 receives the highest overall score and ranks first in both clarity and familiarity but falls short in politeness and ranks third. The yield gesture Y3 comes in second and has an obvious edge in politeness. The Y1 yield gesture ranks third. It scores poorly in the clarity and familiarity categories but has some advantage in politeness. Yielding Gesture Y4 ranked fourth and did not perform well in any of the three dimensions.
\begin{figure}[h]
    \centering
    \includegraphics[width=\linewidth]{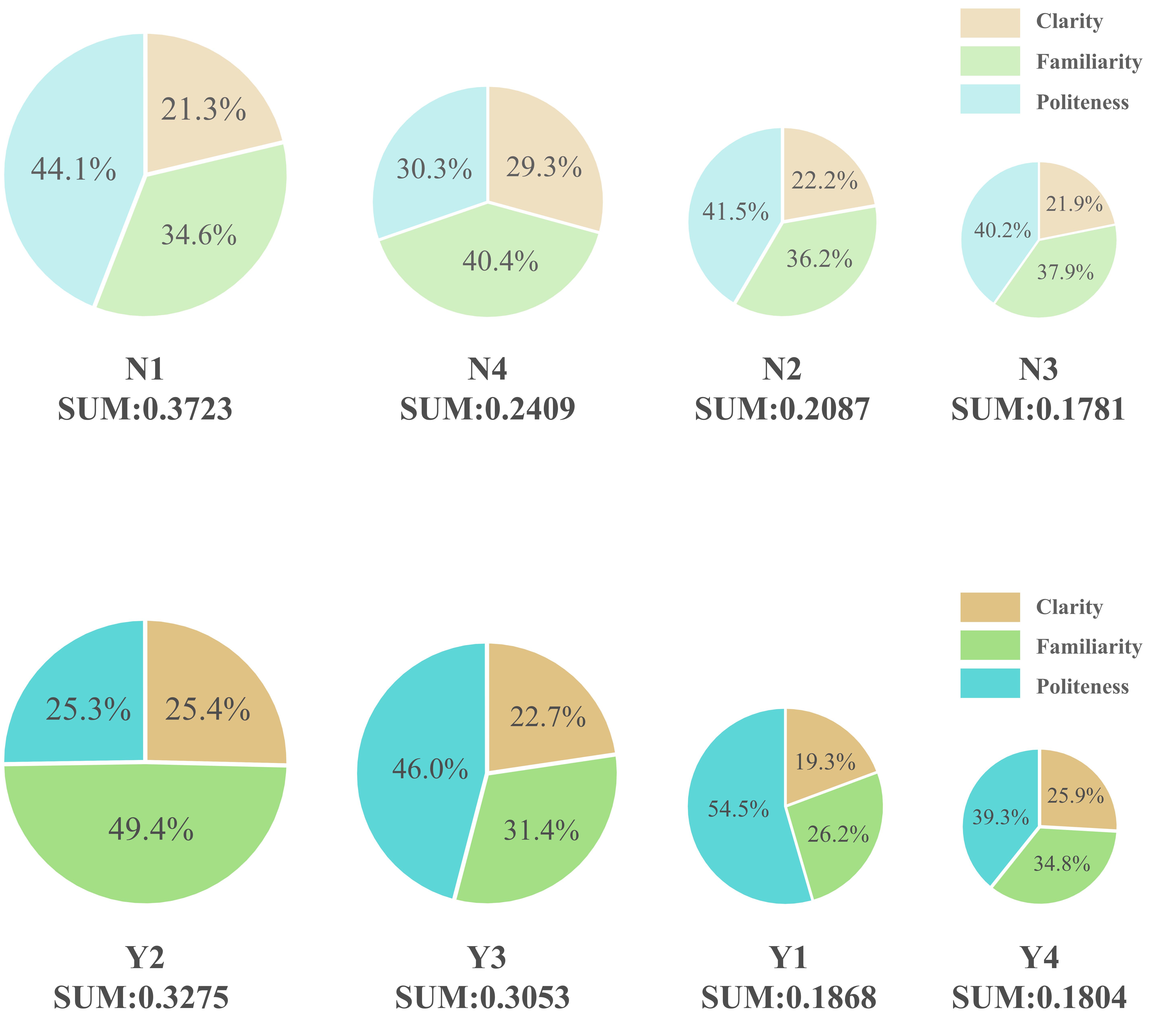}
    \caption{The score of gestures}
    \label{fig:Study2-Score}
\end{figure}
For the non-yielding gestures, the gesture with the highest score was N1 (T-gesture). N4 and N2 received the second and third highest scores, respectively, while N3 was the non-yielding gesture with the lowest total score. The clarity scores for N1 and N4 were higher and closer, and the clarity scores for N2 and N3 were lower and closer, with N1 gesture having the highest clarity score and N3 the lowest. Regarding familiarity, N1 and N4 scored the highest, while N2 and N3 scored lower and similar. Regarding politeness, N1 performed best, while the other three gestures showed little difference in this level. Summarizing the data, the highest total score was obtained by the N1, which performed best in three dimensions, with a more obvious advantage. The non-yielding gesture N4 ranked second, showing better performance in the dimensions of clarity and familiarity but no advantage in politeness. The non-yielding gestures N2 and N3 ranked third and fourth, respectively, and did not perform well in all three dimensions.

\subsubsection{Age differences in three evaluation dimensions}
\
\newline
Other noteworthy phenomena were discovered during the data analysis. When evaluating the importance of the three gesture dimensions, the 12-25 age group emphasized familiarity, while the 25-45 and 45+ age groups emphasized politeness. In the case of yielding gestures, Y2 and Y3 were found to be dominant, with only a small difference between them. The gesture Y3 is more popular among the older age groups because the older age groups place more value on the politeness of the gesture, and Y3 has a clear advantage in terms of politeness. Thus, the phenomenon is observed that younger people prefer Y2 and older people prefer Y3 as the optimal gesture. Opinions on the non-yielding gesture were consistent across age groups, with N1 considered optimal and specific rankings agreed upon.
No statistically significant differences in the gender dimension were found in the study results.
\begin{figure}[h]
    \centering
    \includegraphics[width=\linewidth]{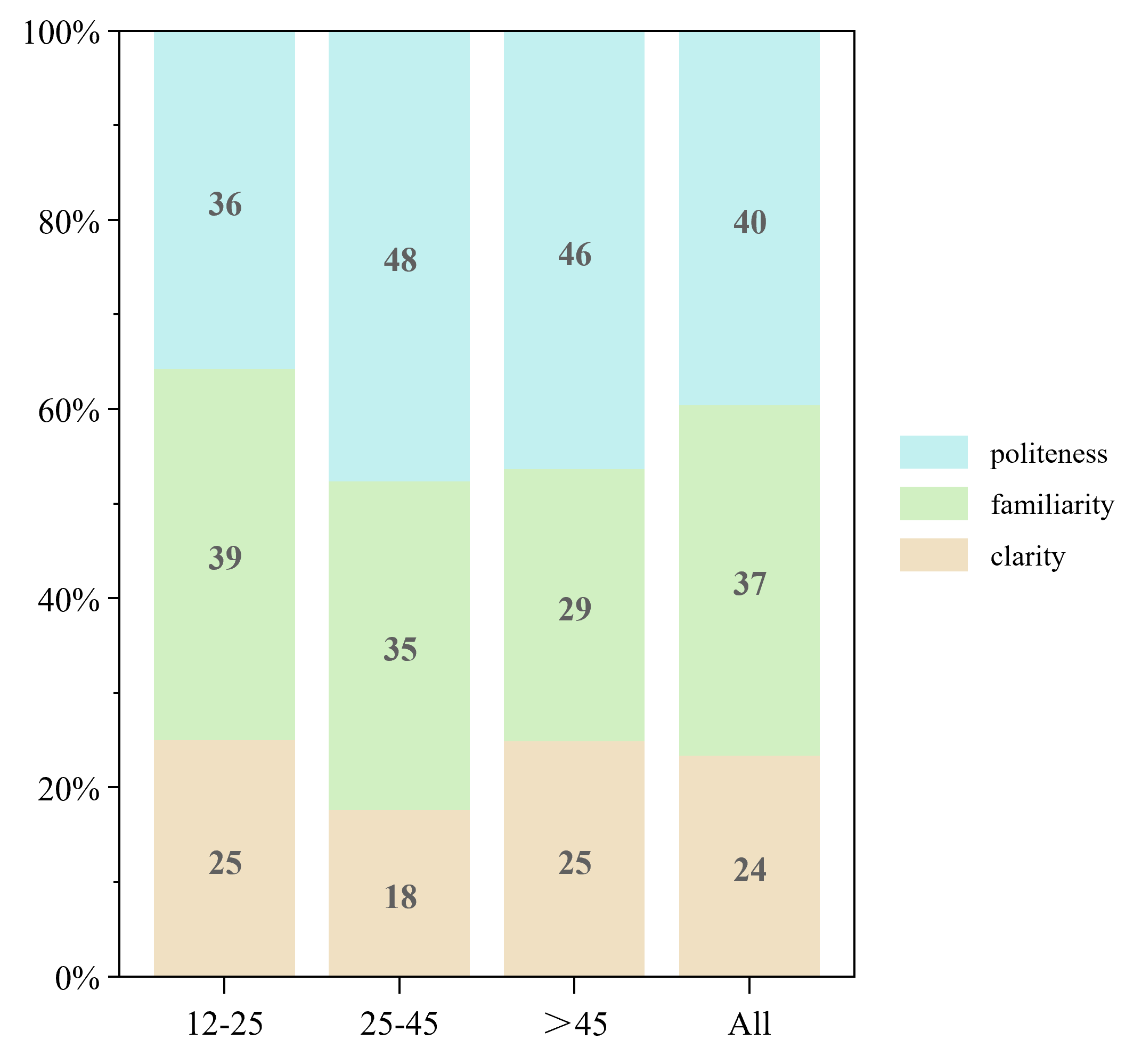}
    \caption{Differences in weight of three dimensions for different ages}
    \label{fig:Study2-age}
\end{figure}
\section{Discussion}

\subsection{Specific receiver and clear content}

Our findings show that the clarity regarding the recipient of the signal is paramount in the interaction between pedestrians and AVs. Among various eHMI methods, gesture-based eHMI stands out as a more effective means of communication. Pedestrians can promptly discern that the AV engages with them through gestures.

Firstly, it is crucial to address the issue of some participants needing help to grasp the role of lighting signals. This could stem from the deeply ingrained habits formed throughout pedestrian-driver interactions since the automobile's invention. Alternatively, the default understanding in traffic scenarios might be that only humans can interpret gestures. Consequently, when a gesture-based eHMI initiates a signal, it instantly captures the attention of other nearby individuals.
In contrast, the lighting of the eHMI lacks clarity for the message receiver. This is due to the ubiquity of lighting applications in contemporary life, resulting in light signals lacking distinct receivers. Lights assume various roles in artificial objects, such as status indicators or decorative elements. Consequently, pedestrians might mistake the car's lights and light bars for mere adornments. Some participants even perceived the light eHMI on the AV as an indicator of the vehicle's status or a means of signaling to other vehicles. In real-world scenarios, a vehicle's lights 10 meters away struggle to draw pedestrians' attention.\cite{app12136730}
Additionally, it is crucial to ensure that pedestrians can perceive the signals the eHMI emits after drawing their attention to them. In the case of gesture-based eHMI, the recognition of actions plays a pivotal role in its effectiveness. The design should make gestures with distinct intentions easily distinguishable, especially from a distance. Beyond a specific range, the eHMI may fail to convey the message, regardless of the gesture. Nevertheless, adjusting factors such as the gesture range, repetition frequency, display position on the car body, and skin color may alleviate this issue. In the future, designers should prioritize these factors when developing gesture-based eHMIs.

\subsection{Overlapping semantics}

The importance of avoiding pedestrian misinterpretation of vehicle intentions stems from the potential overlap in semantics between gestures and lights. In specific scenarios, people associate a particular signal with a specific meaning over an extended period. Consequently, introducing new semantics for the same signal in a different context can lead to confusion and resistance among individuals. Furthermore, the correspondence between colors and intentions only sometimes aligns with common expectations, and participants explained these discrepancies. For instance, in conventional vehicles, red lights have traditionally been used as brake lights, leading people to associate red lights with braking. However, red lights on other artificial objects often signify standby or inactivity. Thus, using red lights in vehicles, eHMI may convey contradictory meanings, potentially baffling participants.
\begin{figure}[h]
    \centering
    \includegraphics[width=\linewidth]{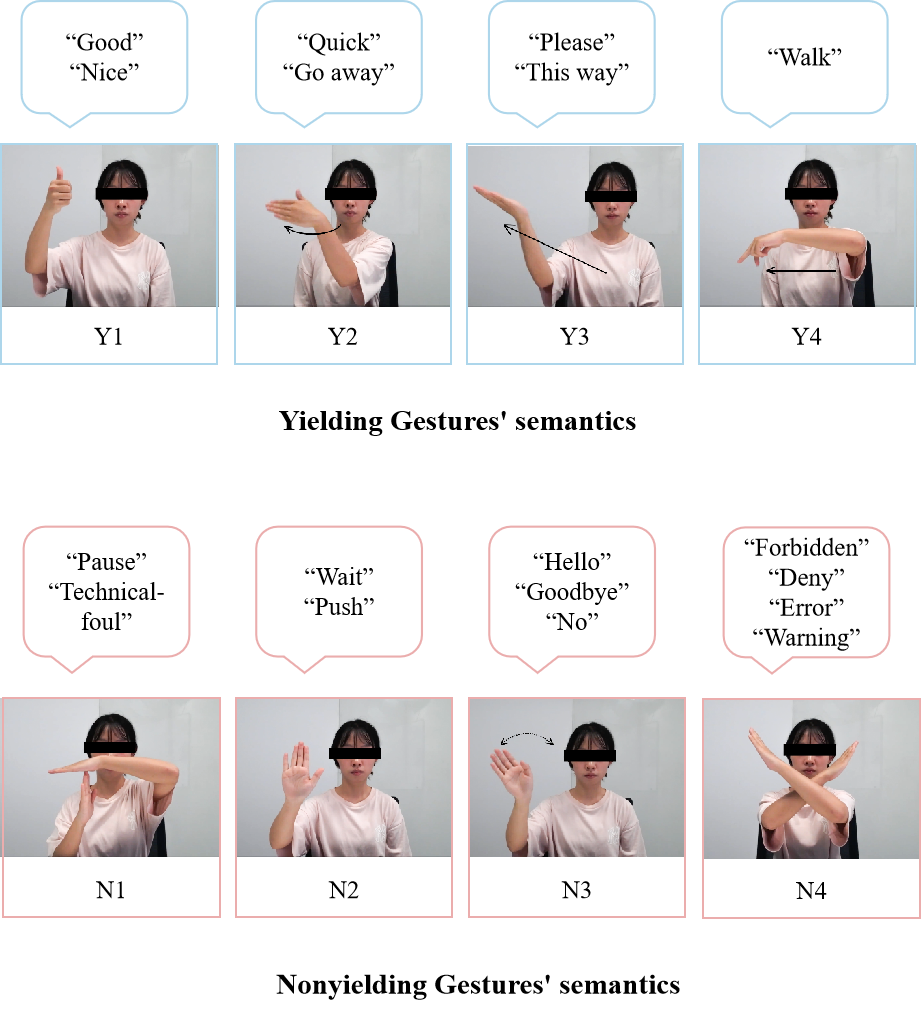}
    \caption{The semantics of gestures}
    \label{fig:Semantics}
\end{figure}

Some gestures exhibit more complex overlapping semantics compared to lights. For instance, Y1 might convey "ok," Y2 could signify "get out of the way," Y3 may denote "please," while N1 could express both "pause" and "technical foul." N3, on the other hand, encompasses meanings such as "hello," "goodbye," and "no," and N4 may convey "reject," "error," "warning," and so on. Notably, not all gestures with overlapping semantics are equally prone to misunderstanding by pedestrians. For gestures like Y2, Y3, N1, and N4, their multiple meanings are often interconnected, making it easier for individuals to relate them to the vehicle's intent in traffic scenarios. This could explain why certain gestures perform as well as, or even better than, light-based eHMI.
However, the overlapping semantics of some gestures can be exceptionally confusing, as seen with the N3 gesture. Drawing parallels between the semantics of "hello," "goodbye," and "do not" is challenging. The meaning of this gesture varies substantially across different scenarios and lacks a fixed interpretation in traffic contexts. Consequently, pedestrians frequently struggle to correctly interpret this gesture, sometimes even more so than an empty vehicle. They tend to misinterpret the intent to yield as the intent not to yield, leading to increased observation and hesitation times when encountering such gestures on two occasions, further exacerbating their confusion. Conversely, gestures like Y4, which are less common in daily life, are quickly understood by pedestrians once they learn their meaning.
Therefore, when designing gestures for vehicle eHMI, it is crucial to consider whether a gesture already carries widely accepted meanings, whether these meanings share commonalities, and whether it might be necessary to create entirely new gestures. Please do so to ensure clarity among pedestrians.

\subsection{A polite gesture or a commanding light?}
Gesture-based eHMI holds a significant advantage over light-based eHMI due to the different modalities, as it can convey a broader range of information. While lights are limited to just two colors, red and green, and offer relatively simple messages, such as "yes" or "no," "go" or "stop," gestures can convey semantic, directional, emotional, and attitudinal information.
Participants who prefer light-based eHMI argue that it is more efficient. They contend that they can quickly discern the car's intentions by simply remembering the meanings of cyan and red lights. They spend more time observing gesture-based eHMI, as extracting visual information from gestures is perceived as a slower process. AV lighting eHMI, with its color information and flashing patterns, allows pedestrians to grasp the situation rapidly. In contrast, processing additional information from gesture-based eHMI takes more time. It is acknowledged that simplicity trumps complexity in this context, making gestures less efficient. Data from Study 1 show that the hesitation time for Y0 and N0 gestures is notably low.
However, it is essential to recognize that pedestrians are not machines but humans with cognitive processes. Combining quantitative and qualitative data, we discover that eHMI not only objectively influences pedestrians' success rates when crossing intersections but also affects their perception of risk. Specifically, gesture-based eHMI tends to make pedestrians feel safer during intersection crossings. Nevertheless, when it comes to the "better" gestures, there is no significant difference in understanding difficulty, observation time, or hesitation time compared to light-based HMI. Study 2 reveals that politeness is crucial, particularly for individuals aged 25-45 and those over 45. The politest gestures, "Y3" and "N1," received high ratings in both studies.
 The utilization of gesture-based eHMI facilitates the interaction between pedestrians and AVs through the involvement of an anthropomorphic "agent", conveying the intentions of AVs to pedestrians, as well as engaging in mutual exchanges with them through gestures. This is similar to the real traffic scenario in daily life. Pedestrians will not consider the action directed towards them as coming from an inanimate device, namely an AV. Instead, they will view the virtual "driver" as a human being who is carrying out interaction and providing responses to the information transmitted by the "agent". Furthermore, gesture-based eHMI will enhance the safety of pedestrians and gain their trust due to the trust people have in real drivers. 

 \subsection{Learning during experiencing}

 When comparing pedestrians' performance in encounters with different eHMIs over three instances, we discern some differences in their interactions with autonomous vehicles equipped with various eHMIs. Participants tended to learn during experiencing. Different eHMIs exhibit varying learning costs, manifesting in distinct performance levels during initial encounters. Moreover, there are discernible differences in learning rates, with pedestrian performance improving as the number of encounters increases.
 
Regarding learning costs, the study suggests that the more familiar an eHMI is in daily life, the lower the associated learning cost~\cite{Guo_Yuan_Yu_Chen_Yu_Cheng_Wang_Luo_Jiang_2022}. Pedestrians perform better during their initial encounter with eHMIs resembling familiar displays. Light-based eHMI, a relatively standard and recognizable format, yields shorter hesitation times when it appears for the first time, as exemplified by "Y0." Its resemblance to everyday traffic lights contributes to this familiarity. Similarly, when pedestrians encounter an unfamiliar gesture-based eHMI for the first time, gestures commonly used in daily life incur lower learning costs and are grasped more swiftly. For instance, in Study 1, pedestrians exhibited the shortest observation time when first encountering "Y2," a gesture frequently used in daily life. The findings in Study 2 corroborate this observation, particularly with "Y2" performing exceptionally well in a survey based on a questionnaire, which entails only one repetition.

Concerning learning rates, it is evident that different eHMIs exhibit distinct tendencies. Light-based eHMI mainly conveys intentions through colors and logo animations, fostering stable learning and memory of abstract graphics associated with vehicle intentions. For instance, the decrease in understanding difficulty and hesitation time for "Y0" follows a nearly linear trend. Even unusual gestures have the potential to be rapidly learned and accepted. Take "Y4" as an example: despite its initial lack of familiarity, it impressed pedestrians due to its straightforward and easily recognizable characteristics, coupled with instructions that lacked overlapping semantics with the direction of pedestrian movement.
As a result, "Y4" exhibited that it is learned the best by participants, performing exceptionally well by the third encounter.

\subsection{Future work and limitations}

This research has several limitations. Firstly, our VR experiments focused only on the interaction between autonomous vehicles and pedestrians, excluding other traffic participants. Future research needs to investigate the interaction effects of autonomous vehicles using gesture-based eHMI in more complex mixed traffic scenarios, involving pedestrians, cyclists, and other non-autonomous vehicles. Additionally, one of the challenges in the development of autonomous vehicles is the "open set" nature of the environment, with countless factors that could pose challenges. Pedestrian behavior varies widely, and future work should explore whether different factors, such as different speeds, day or night conditions, single or multiple pedestrians, impact the interaction with autonomous vehicles.

Furthermore, we were interested in combining lighting and gestures to see if there would be better results. However, due to the eight different gestures used, combining them with lighting or other types of eHMI would result in a large number of experimental conditions. This would make the participation time in VR experiments very long and could easily lead to dizziness or nausea. Attempting to combine gestures and lighting and testing their effects is undoubtedly a promising avenue for future research.

Similarly, during the internal testing phase of the experiment, we attempted to have participants stand and wear VR headsets, physically walking in the real world to control the motion of their virtual pedestrian avatar. However, we found this to be severely limited by physical space and hardware constraints, leading to significant discomfort and difficulty for participants to complete the entire experiment. Therefore, we ultimately modified the experiment design to have participants sit and observe approaching autonomous vehicles, using buttons to control their pedestrian avatar. We raised the initial height of the first-person camera to ensure consistency between the seated perspective and the normal walking perspective. In future research, the use of VR walking simulators, such as flat-based or bowl-based omnidirectional treadmills~\cite{Hooks_Ferguson_Morillo_Cruz-Neira_2020}, may yield better results to ensure immersion and measure changes in pedestrian walking speed and direction.

Secondly, we were unable to address the issue of gesture-based interaction interfaces being difficult to identify and understand when the vehicle is far from the pedestrian. In VR environments, due to issues such as resolution and texture, it is challenging for the human eye to discern signals from several tens of meters away. This problem may improve in the real world, but it remains a focal point for future work. If interactions between pedestrians and autonomous vehicles can occur at greater distances, it would provide more space and reaction time for all traffic participants.

Since this study was conducted only in China, some of the chosen gestures may not be applicable in other countries or regions. Future research could further explore cultural differences and broaden the scope of experiments. In the survey study, we only selected age and gender as common indicators to analyze preference differences among different demographics. However, education, driving qualifications, and other indicators could also lead to preference differences. Due to the questionnaire of study 2 being distributed on internet platforms, participants may tend to be relatively young, and the gesture score may not represent the preferences of a more diverse population. Nevertheless, our investigation into age differences provides insightful findings. Future work could explore the impact of different gestures in a more diverse population.

In future work, to further validate the effectiveness of gesture-based eHMI, it is inevitable to work within actual vehicle systems. Nowadays, automotive manufacturers are already experimenting with augmented reality projections to display content on car windows, achieving high visibility, as demonstrated by the BMW i Vision Dee at CES 2023~\footnote{https://www.bmwgroup.com/en/news/general/2023/i-vision-dee.html}. However, accepting gestures that prohibit crossing for many accustomed to crossing in front of vehicles may be challenging. It will take time for people to adapt to observing signals conveyed by the eHMI of AVs at intersections to understand the vehicle's intent and take action. Of course, this also requires cooperation with local policies and laws. Nevertheless, this is undoubtedly a trend in the future development of AVs, providing a means for AVs to communicate with pedestrians.

\section{CONCLUSION}

Eight eHMI were designed based on common gestures to convey an autonomous vehicle’s intention to yield or not yield at an uncontrolled crosswalk. Through VR experiments (N1 = 31) and online surveys (N2 = 394), we found significant differences in usability between gesture-based eHMIs compared to other eHMIs.We explored the reasons for the differences in eHMI effects. The signal recipient is not clear, which will result in confusion. Whether the gestures are clear, polite, and familiar to pedestrians is taken into consideration by pedestrians. We encourage designers and researchers to support interactions between autonomous vehicles and other road users through understandable gesture-based eHMI and suggest potential opportunities and future research directions.

\begin{acks}
This project is supported by the Beijing Natural Science Foundation (Grant No.L233033). We thank the participants who shared their lived experiences with us.
\end{acks}

\bibliographystyle{ACM-Reference-Format}
\bibliography{main}

\clearpage
\begin{appendix}
\onecolumn
\section{Supplementary Note to Study 2}\label{appendix:Supplementary Note to Study 2}
\subsection{Introduction}
This document provides further explanation of the design and analysis concerns in the STUDY2, questionnaire research section.

\subsection{Questionnaire Structure}
The questionnaire initially obtained basic demographic information from participants, including age and gender. Subsequently, the questionnaire classified the factors for evaluating gestures into clarity, familiarity, and politeness. The questionnaire requested participants to assess these three dimensions of evaluating gestures by comparing them in two-by-two scenarios and by rating various gestures on the three dimensions (clarity, familiarity, and politeness) in two-by-two comparisons~\cite{forman2001analytic,saaty2008decision,vaidya2006analytic}.

\begin{figure}[htb]
    \centering
    \includegraphics[width=0.35\linewidth]{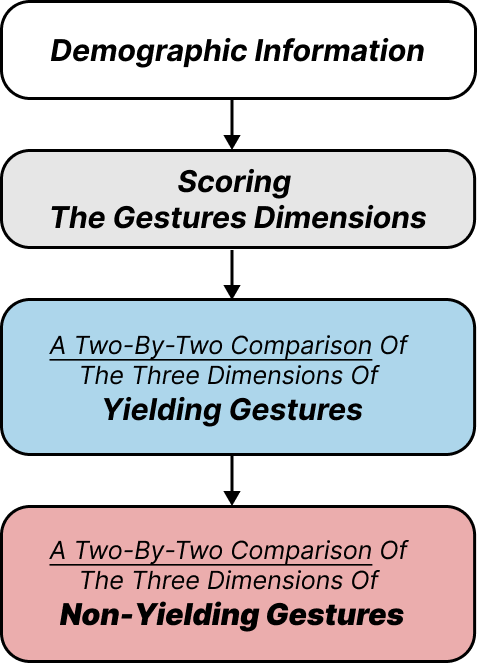}
    \caption{Questionaire Structure}
    \label{fig:Questionaire Structure}
\end{figure}

The questionnaire participants were given clarification on the three dimensions beforehand. Clarity, which measures the clarity of the gesture in expressing intention. Familiarity, which measures the degree of familiarity with gestures in everyday life. Politeness, which measures the politeness of the gesture in expressing intention.

\subsection{Questionnaire Design}
\subsubsection{Data Quality}
\ 
\newline
\indent Several measures were implemented to ensure data quality in this questionnaire study. The questions were written in simple, understandable language to aid comprehension, and the questionnaire included motion pictures with gestures to improve visual understanding.
Additionally, participants were informed that the data collection was completely anonymous to avoid bias. Furthermore, two questions were included in the survey to eliminate the possibility of machine responses. Lastly, any questionnaire that was completed too hastily or took an excessive amount of time was deemed invalid.

\subsubsection{Topic Design and Analysis Methods}
\ 
\newline
\indent 
Analytic Hierarchy Process (AHP) is a renowned method for decision-making, initially proposed by American operations researcher, Saaty, in the early 1970s. It involves decomposing decision-making elements into levels such as objectives, criteria, and options, which are then qualitatively and quantitatively analyzed through pairwise comparisons. 
The initial step involves structuring the problem into a hierarchy (Figure 21). The highest level entails selecting the optimal yielding and non-yielding gesture. The second level comprises three criteria that contribute to achieving this goal. The third level consists of eight gestures that need to be evaluated considering the criteria of the second level.

\begin{figure}[htb]
    \centering
    \includegraphics[width=0.75\linewidth]{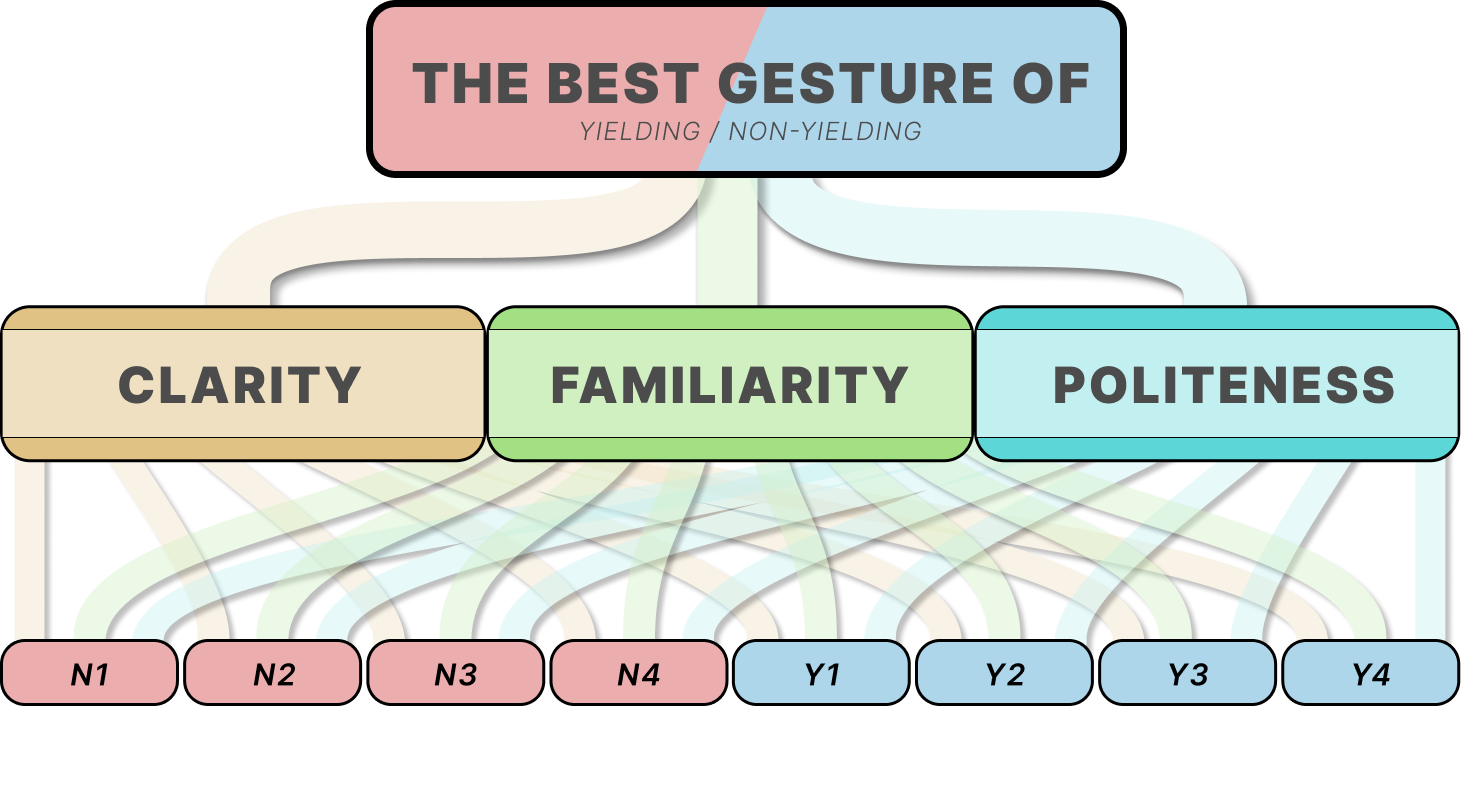}
    \caption{The hirarchy model of gesture}
    \label{fig:The hirarchy model of gesture}
\end{figure}

The second step is to construct a comparative judgment matrix. A comparative rating matrix is created by comparing factors two-by-two, using the criteria from the previous level. $A_{ij}$ may be utilized to express the relative importance of the ith and jth elements, with a typical value range of 1-9 and their reciprocal. The resulting $A_{ij}$ values constitute the judgment matrix, which is also known as the comparison matrix. For instance, the second layer of the three evaluation dimensions will be compared in pairs to obtain the relative importance of the three dimensions. Additionally, the gestures are compared in pairs again, this time under the clarity dimension, to determine the relative importance of each gesture under that dimension. To make it easier to understand, each comparison is designed as a slider input.

\begin{figure}[htb]
    \centering
    \includegraphics[width=0.75\linewidth]{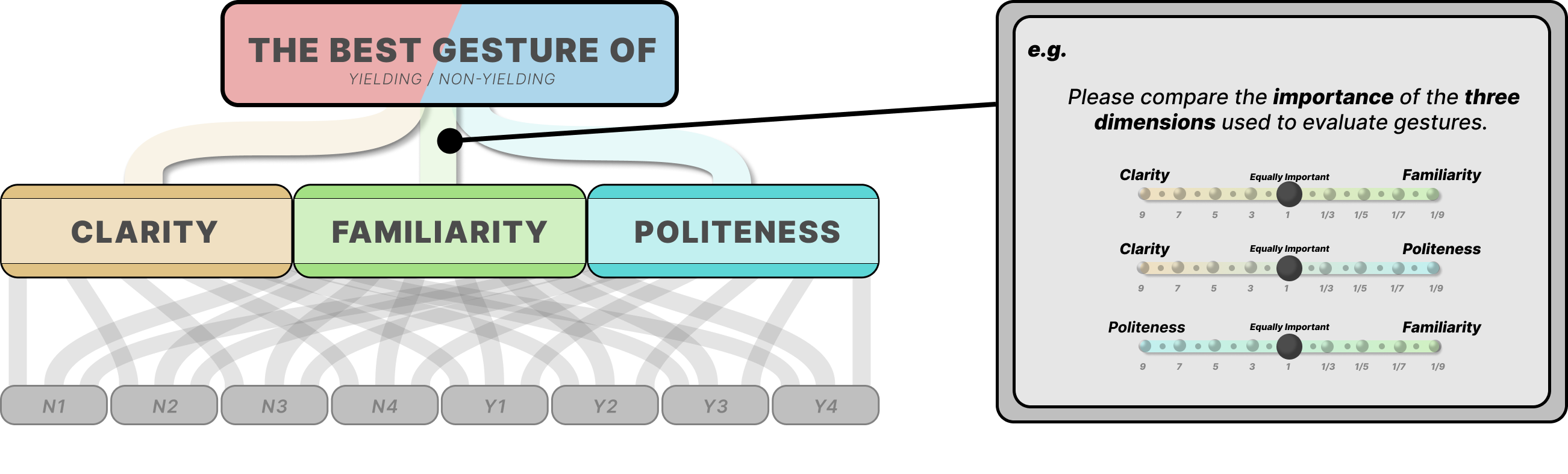}
    \caption{STEP2-1 Dimension Comparison}
    \label{fig:STEP2-1 Dimension Comparison}
\end{figure}
\begin{figure}[htb]
    \centering
    \includegraphics[width=0.75\linewidth]{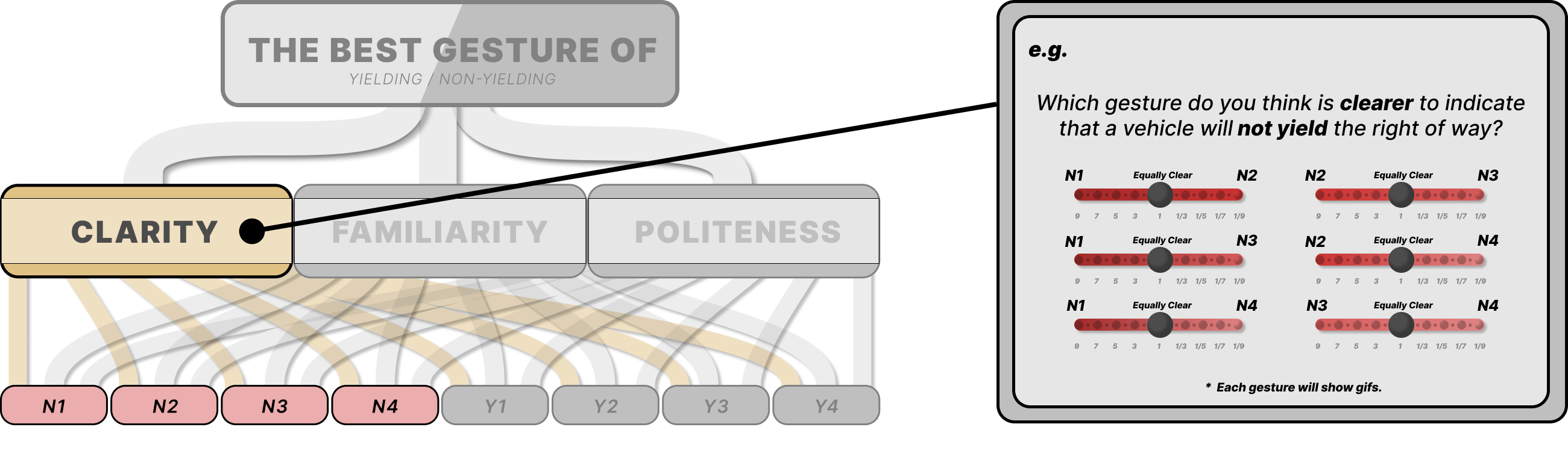}
    \caption{STEP2-2 Gesture Comparison}
    \label{fig:STEP2-2 Gesture Comparison}
\end{figure}

The third step involves conducting hierarchical single ranking and its consistency test. By obtaining a characteristic equation for each comparison judgment matrix, the equation can then be solved to give the solution vector and regularized, forming an importance ranking of factors at the same level with comparable factors at the previous level. For example, after determining the relative importance of the three dimensions at the criterion level, the judgment matrix is processed to obtain the final solution vector which reflects the weight share of each dimension for selecting the optimal gesture. The process is then repeated to determine the score of each gesture in terms of clarity within the clarity dimension.

The fourth step involves conducting a hierarchical total ordering and a consistency test. This requires calculating the relative importance of all elements within the same level to the target level. The resulting hierarchical total ranking must be tested for consistency to guarantee the reliability of the findings. After completing the aforementioned steps, the calculation process yields the weights of the three dimensions, scores for each gesture in each dimension, and, ultimately, the overall score for each gesture. Prior to score calculation, it becomes necessary to unify the group opinion by processing the geometric mean to establish a unified judgment matrix.

\begin{figure}[htb]
    \centering
    \includegraphics[width=0.75\linewidth]{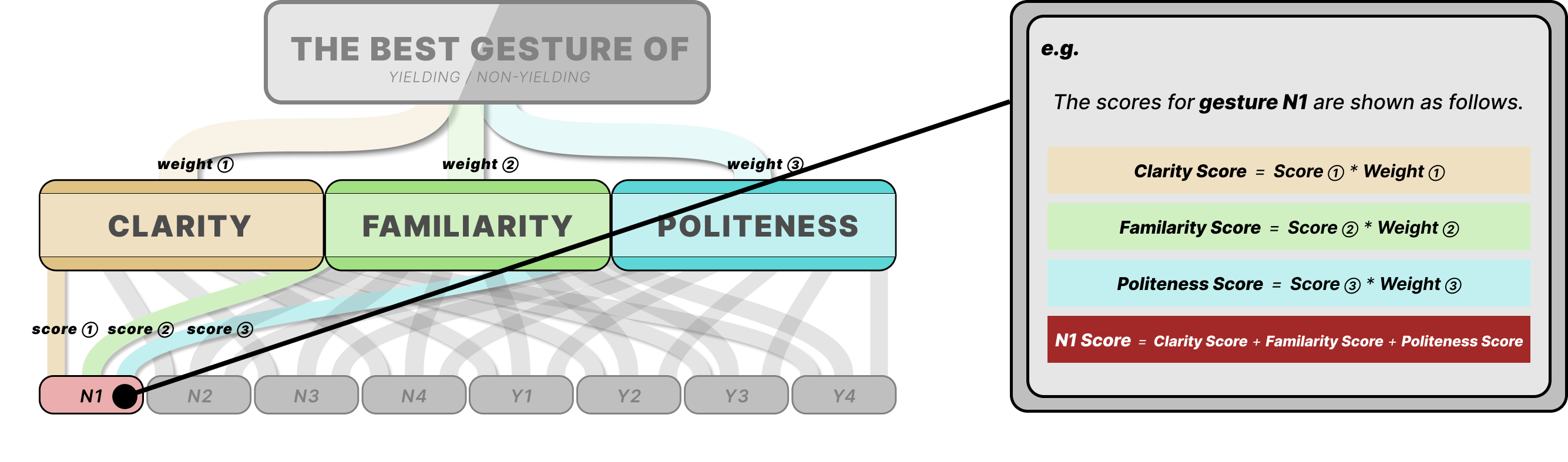}
    \caption{STEP3+4 Gesture Scoring}
    \label{fig:STEP3+4 Gesture Scoring}
\end{figure}

\subsection{Questionnaire Design}
The survey was formulated using the AHP analysis technique, resulting in questions being presented in a two-by-two comparison format. This approach obtains the assessors' assessment of the significance of the three dimensions and the effectiveness of each gesture in each dimension, allowing the researchers to achieve a more exhaustive analysis of the gestures. This approach obtains the assessors' assessment of the significance of the three dimensions and the effectiveness of each gesture in each dimension, allowing the researchers to achieve a more thorough analysis of the gestures.
\end{appendix}

\end{document}